\documentclass[fleqn,usenatbib,twocolumn]{aastex63}

\usepackage[T1]{fontenc}
\usepackage{ae,aecompl}

\usepackage{graphicx}	
\usepackage{amsmath}	
\DeclareMathOperator{\sech}{sech}

\usepackage{amssymb}	

\usepackage{txfonts} 

\usepackage{epstopdf}

\usepackage[none]{hyphenat}
\graphicspath{{images_aj_formatted/}}
\DeclareGraphicsExtensions{.pdf,.png,.jpg, .jpeg}
\usepackage{natbib}
\usepackage{color}

\usepackage{threeparttable}

\shorttitle{Orbital ingredients for cooking X-structures}
\shortauthors{Parul et al.}
 
\begin{document}
\title[Orbital ingredients for cooking X-structures]{Orbital ingredients for cooking X-structures in edge-on galaxies}

\correspondingauthor{Natalia Ya. Sotnikova}
\email{n.sotnikova@spbu.ru}

\author{Hanna D. Parul}
\affiliation{Department of Physics and Astronomy, University of Alabama, P.O. Box 870324, Tuscaloosa, AL
35487-0324, USA}
\affiliation{St. Petersburg State University,
Universitetskij pr.~28, 198504 St. Petersburg, Stary Peterhof, Russia}

\author{Anton A. Smirnov}
\affiliation{St. Petersburg State University,
Universitetskij pr.~28, 198504 St. Petersburg, Stary Peterhof, Russia}
\affiliation{Central Astronomical Observatory at Pulkovo of RAS, Pulkovskoye Chaussee 65/1, 196140 St. Petersburg, Russia}

\author{Natalia Ya. Sotnikova}
\affiliation{St. Petersburg State University,
Universitetskij pr.~28, 198504 St. Petersburg, Stary Peterhof, Russia}

\received{June 1, 2019}
\revised{January 10, 2019}
\accepted{\today}

\submitjournal{ApJ}

\begin{abstract}
X-structures are often observed in galaxies hosting the so-called B/PS (boxy/peanuts) bulges and \textcolor{black}{are} visible from the edge-on view. They are the most notable features of B/PS bulges and appear as \textcolor{black}{four rays protruding from the disk of the host galaxy and distinguishable against the B/PS bulge background.}
\textcolor{black}{In some works their origin} \textcolor{black}{is} thought to be connected with the so-called banana-shaped orbits with a vertical resonance 2:1. A star in such an orbit performs two oscillations in the vertical direction per one 
\textcolor{black}{revolution in the bar frame.}
Several recent studies that analyzed ensembles of orbits arising in different \textcolor{black}{$N$-body models} do not confirm the dominance of the resonant 2:1 orbits in X-structures. In our work we \textcolor{black}{analyze two $N$-body models and} show how the X-structure \textcolor{black}{in our models} is gradually assembled from the center to the periphery from orbits with less than 2:1 frequency \textcolor{black}{ratio}. \textcolor{black}{The most number of such orbits is enclosed in a `farfalle'-shape (Italian pasta) form and turns out to be non-periodic.}
We conclude that \textcolor{black}{the X-structure is} only the envelope \textcolor{black}{of regions of high density caused by the crossing or folding of different types of orbits at their highest points,
and does not} have a ``backbone'' similar to that \textcolor{black}{of} the in-plane bar. Comparing the orbital structure of two different numerical models, we show that the dominance of one or another family of orbits with a certain ratio of the vertical oscillations frequency to the in-plane frequency depends
on the parameters of the underlying galaxy and ultimately determines the morphology of the X-structure and the opening angle of its rays.  
\end{abstract}


\section{Introduction}
Morphological studies of S0 -- Sd galaxies in close to edge-on orientation have revealed that a significant fraction of them, up to 45\%, possess peculiar boxy or peanut-shaped (B/PS) isophotes in central regions \citep{Lutticke_etal2000}. 
\par
B/PS morphology is found to arise naturally in simulations of a secular growth of a bar in the vertical direction. 
\citet{Combes_Sanders1981} were the first who demonstrated that such a 3D bar seen edge-on looks like a box when viewed with the line of sight parallel to the long axis of the bar. When the line of sight is perpendicular to the long axis of the bar (a side-on view), it exhibits a peanut-shape morphology. \textcolor{black}{Such a morphology is distinguishable in a certain range of the position angle (PA) of the bar. The strict range is difficult to determine, since it depends on the resolution of a particular image and properties of the B/PS bulge itself.} Further numerical studies \citep{Combes_etal1990,Pfenniger_Friedli1991,Raha_etal1991,Athanassoula_Misiriotis2002,Oneil_Dubinski2003,Athanassoula2005,MartinezValpuesta_etal2006,Debattista_etal2006} showed that observed B/PS bulges can be associated with a bar that has grown in the vertical direction.
\par
One of the most prominent features of B/PS bulges is a bright cross \textcolor{black}{sticking out of} the image of the host galaxy seen edge-on (see, for example, \citealp{Bureau_etal2006}).
 The cross is called an X-structure and is almost \textcolor{black}{symmetric} with respect to a mid-plane of the stellar disk and the plane passing through the center of the disk, perpendicular to the mid-plane of the stellar disk.
\par
Historically, the first galaxy, \textcolor{black}{which revealed} the X-structure, was the peculiar S0 galaxy IC~4767 \citep{Whitmore_Bell1988}. \citet{Whitmore_Bell1988} enhanced the X-structure by subtracting the bulge and disk models from the image. The second galaxy was the galaxy Hickson 87a of Sbc type \citep{Mihos_etal1995}. The X-structure of this galaxy can be seen in the unsharp-masked image, although in the HST image \citep{English_etal2000} it is visible simply by the eye.
\par
Initially, X-structures found in $N$-body simulations were considered as an illusory effect \citep{Friedli_Pfenniger1990,Pfenniger_Friedli1991}, and not as specific bright details of the B/PS bulges, but later they were massively identified in $N$-body snapshots processed by unsharp masking procedure \citep{Athanassoula2005}. In models with a high resolution, an X-structure is directly distinguished as two diagonals of the boxy structure in the gray scale pictures without any processing \citep{Laurikainen_etal2014,Salo_Laurikainen2017,Smirnov_Sotnikova2018}. In the observational data, the X-structures are found either by unsharp masking procedure \citep{Bureau_etal2006}, or by analyzing the sixth Fourier mode of edge-on galaxy isophotes \citep{Ciambur_Graham2016}, or by subtracting the disk and bulge models from the image of the galaxy \citep{Savchenko_etal2017}.
\par
The X-structures are local enhancements within B/PS bulges which nature is tightly connected with the origin of peanut-like bars. Two possible mechanisms contribute to the formation of B/PS bulges in disk galaxies. \citet{Raha_etal1991} suggested that such a structure can occur due to the violent buckling instability of a bar similar to the global bending instability of a fire-hose type of a stellar disk \citep{Toomre1966,Poliachenko_Shukhman1977,Araki1985,Merritt_Sellwood1994}. 
The buckling corrupts the bar's midplane and affects mostly the central part of a disk where the rotation curve is rising \citep{Merritt_Sellwood1994,Sellwood1996,Sotnikova_Rodionov2003,Sotnikova_Rodionov2005,MartinezValpuesta_etal2004,MartinezValpuesta_etal2006}. Another mechanism is a gradual vertically symmetric growth of a bar via resonant trapping of individual stars \citep{Quillen2002,Quillen_etal2014}. In this case a growing bar is populated \textcolor{black}{by 3D orbits bifurcated from the so-called x1 family of orbits} (in notation by \citealp{Contopoulos_Papayannopoulos1980}), which are elongated along the 2D bar and support it. The dominant x1 orbit family may become vertically unstable and, as the bar grows, 3D orbits with higher and higher vertical resonances are involved in it. 
\par
\textcolor{black}{
There is a well-developed area of nonlinear dynamics 
in which 3D periodic, quasi-periodic (q.p.) and different types of chaotic orbits are studied. Some of these orbits can constitute the backbone of B/PS bulges and can delineate X-structures \citep{Pfenniger_1984,Pfenniger_Friedli1991,Skokos_etal2002,Patsis_etal2002a,Patsis_etal2002b,Patsis_Katsanikas2014a,Patsis_Katsanikas2014b,Patsis_Harsoula2018,Patsis_Athanassoula2019}. Usually, periodic and quasi-periodic orbits are calculated in analytic potentials representing a disk, a bar, and a bulge. The results of many mentioned works have become classic and are often used to interpret the observational data. In some works orbits from the x1v1 (following the notation by \citealp{Skokos_etal2002}), or BAN (in the notation by \citealp{Pfenniger_Friedli1991}) family, which appears at the vertical 2:1 resonance (i.e. 2 vertical oscillations per one revolution in the bar frame), were considered as the main building blocks of the B/PS shape. Within this approach, the X-structure is a sharp edge of the `peanut' produced by the successive alignment of $z$-maxima of x1v1 orbits \citep{Patsis_etal2002b,Patsis_Katsanikas2014a}, which are often called `banana' orbits because of their banana shape when seen side-on. Higher order members of the x1 tree (3:1 and even 4:1) can also contribute to the peanut shape in considered potentials at large distances from the center \citep{Skokos_etal2002,Patsis_etal2002b}. Another family of orbits with an infinity symbol profile (x1v2, or ABAN), bifurcating from x1 at the 2:1 vertical resonance, can also contribute to the X-structure \citep{Patsis_Harsoula2018}.}
\par
\textcolor{black}{\citet{Portail_etal2015b} doubted such a scheme. They analyzed the orbital structure of a set of $N$-body models of barred galaxies. Their models included an exponential disk and a Navarro-Frenk-White (NFW) halo \citep{NFW}. \citet{Portail_etal2015b} argued that 3D periodic orbits associated with 2:1 vertical resonance cannot explain the vertical X-shape of a bar in their models because the orbits associated with this resonance are few and appear primarily in the outermost regions. Instead they proposed that resonant boxlet orbits associated with the 5:3 vertical resonance (the so-called `brezel' orbits) are responsible for the X-shape in the inner parts of a bar. Moreover, \citet{Portail_etal2015b} did not found any members of the x1 tree higher than 2:1 in the vertical structure of a bar.} 
\par
\textcolor{black}{Orbital composition of B/PS bulges was investigated in some other works, where $N$-body models were considered. 
\citet{MartinezValpuesta_etal2006} used characteristic diagrams for their $N$-body snapshots to study the orbital families and their stability.
\citet{Wozniak_Michel-Dansac2009,Valluri_etal2016,Abbott_etal2017} studied orbits in their $N$-body simulations of an isolated disk galaxy using the frozen potential method as \citet{Portail_etal2015b}, while \citet{Gajda_2016} traced the orbits directly in an $N$-body potential of a dwarf galaxy with a tidally induced bar.
\citet{Valluri_etal2016} examined two $N$-body simulations of self-consistent bars, which arise in potentials of an exponential disk and a NFW halo, and showed that all families of orbits found in triaxial ellipsoids (for example, \citealp{deZeeuw_1985}) are present in their $N$-body bars. They evaluated the fractions of orbits associated with these different families. For their models, the most populated family was found to be the so-called `box' family of orbits, not x1. Further \citet{Abbott_etal2017} using the same models investigated how these families participate in building the X-structure in the galaxy of the Milky Way type. The results by \citet{Abbott_etal2017} are in broad agreement with the findings of \citet{Portail_etal2015b}. First of all, this calls into question the role of the resonant 2:1 orbits in the formation of X-structures. In models by \citet{Valluri_etal2016} and \citet{Abbott_etal2017} such orbits are few, comprise only about 3\% of all bar orbits and they are found primarily in the outer half of the bar. Thus these families are not capable of reproducing all spatial and kinematic characteristics associated with the X-shape in models by \citet{Valluri_etal2016} and \citet{Abbott_etal2017}. Moreover, these orbits combined together show shell-like structures in unsharp masked images, rather than X-structures. Although \citet{Gajda_2016} found a significant number of (A)BAN orbits, their model has many orbits with different vertical morphology, especially in the inner part of the bar (their figure~10). On the contrary, \citet{Wozniak_Michel-Dansac2009} found a large number of orbits with the vertical resonance of 2:1.}
\par
Contradictory results concerning the orbital composition of B/PS bulges and X-structures can be reconciled, since the models under consideration are very different. \textcolor{black}{Works in the field of nonlinear dynamics most often deal with the \cite{Miyamoto_Nagai1975} disk. A triaxial Ferrer's ellipsoid representing a bar and a small bulge are superimposed on this model. In such a combined potential, the most 
studied family of 3D periodic orbits is the x1v1 family. In contrast, quite other potentials are used in $N$-body simulations. The models consist of an exponential disk, a NFW halo, and the bar, which was formed in a self-consistent manner. Although such models may vary in detail, \textcolor{black}{they all demonstrate only a few of x1v1 orbits.}\footnote{Two particular works that dealt with $N$-body potentials stand apart for the following reasons. \citet{Wozniak_Michel-Dansac2009} operate with a Miyamoto-Nagai disk in their $N$-body simulations. An $N$-body model by \citet{MartinezValpuesta_etal2006} consisted of an exponential disk and a halo, but the authors did not study in detail the orbital composition of their B/PS bulge and only examined the role of x1v1 and x1v2 orbits in the X-shape formation.}. 
\textcolor{black}{Recently, the origin of `non-x1-tree', bar-supporting orbits, the so-called mul-periodic orbits, has been studied in \citet{Patsis_Athanassoula2019}. Most of these orbits can be associated 
with resonant boxlet orbits from $N$-body simulations and they have a side-on shape, different from a `banana' or an infinity sign.}
In addition}, in a series of $N$-body simulations \citet{Smirnov_Sotnikova2018} showed that the morphology of the X-structure and some of its parameters depend on parameters of the underlying galaxy. In particular, the authors showed that the opening angle of the X-structures (the angle between the ray and the major axis of the galaxy), the extension of the rays and their brightness strongly depend on the dark halo contribution to the overall gravitational potential. The Toomre parameter $Q$ \citep{Toomre1964} can also influence the morphology of the resulting X-shape. Some simulated galaxies in \citet{Smirnov_Sotnikova2018}, which were dynamically hot enough in disk plane, demonstrated even a double X-shape with a secondary X-structure (in the sense of the birth time), which was off-centered, more flattened and longer than one in the center of the disk.
\par 
Observational data also \textcolor{black}{show} the diversity of X-structure types. The opening angles of the observed X-structures range from $20^{\circ}$ to $45^{\circ}$ \citep{Ciambur_Graham2016,Laurikainen_Salo2017,Savchenko_etal2017}. 
\textcolor{black}{\citet{Savchenko_etal2017} demonstrated that the projection effects cannot fully explain the observed spread in the opening angles of X-structures, although the distribution over the opening angles is contaminated by these effects when a bar is viewed from different PAs.} In addition, \citet{Bureau_etal2006} \textcolor{black}{noted} that X-shapes come in two types, centered and off-centered.
\par
Clearly, the observed morphology of X-structures should be related to the orbital composition of the X-structure \textcolor{black}{itself}. 
\textcolor{black}{And the main question is not which orbits supporting the B/PS or X-like shape can exist theoretically, but which of them specifically manifest (dominate) in one or another potential.}
In the present work, we suggest that the diversity of the X-structures is due to the dominance of certain \textcolor{black}{near periodic orbits.}
\par
To test this we analyzed actual particle orbits in the ongoing potential of two $N$-body models.
In most studies performed to date the orbits were analyzed in either an analytic or in so-called frozen potential, which is extracted from $N$-body simulations well after the formation and buckling of the bar. We use another approach as in \citet{Ceverino_Klypin2007}, \citet{Wang_etal2016} and \citet{Gajda_2016}. We decided to take this more \textcolor{black}{cumbersome} approach for two reasons. First, there are still too few studies where orbits in the evolving potential of the $N$-body model were analyzed. Secondly, this approach is easier to extend for other problems of the orbital dynamics in the future. For example, it can be used for studying the evolution of orbits over a long period of time, comparable with the characteristic time of the global evolution of models. The first attempt to analyze how the distribution of the frequency ratios changes over a short period of time as the bar \textcolor{black}{grows} in the vertical direction was made by \citet{Lokas2019}. Such studies need further and extensive development. 
\par
We re-calculated and re-analyzed two $N$-body multi-component models with the same relative mass of the dark matter and different values of the Toomre parameter $Q$ from \citet{Smirnov_Sotnikova2018}. For these models the resulting values of the opening angle are different. Moreover, the hotter model has a double X-structure. We carried out a frequency analysis of all the orbits involved in the bar, using the method described in \citet{Gajda_2016}, and showed how the X-structure is assembled from different  \textcolor{black}{types of} orbits throughout its entire length.
\par
The rest of the paper is organized as follows.
In Section~2 we give an overall description of numerical models and characterize the simulation of barred disks used.
In Section~3 we describe the details of the frequency analysis method.
In Section~4, we show how the particles of a bar and a disk are separated.
In Section~5, we analyze the distribution of frequency ratios in many aspects and study the spatial distribution of different groups of orbits.
In Section~6, we reveal the orbital anatomy of B/PS bulges.  
In Section~7, we study individual typical orbits and their possible contribution to the assembly of X-structures.
In Section~8, we give an interpretation of the phenomenon of X-structures. 
Finally, in Section~9 we summarize our results and compare them with previous works.

\section{Methods}

\subsection{Numerical models}
We re-considered two three-dimensional $N$-body models of galaxies from \citet{Smirnov_Sotnikova2018}. Each model comprises a stellar disk and a dark spherical halo.
\par
The disk has the total mass $M_\mathrm{d}$ and its density profile is exponential in a radial direction with a scale length $R_\mathrm{d}$ and consists of isothermal sheets in the vertical direction with a scale height $z_\mathrm{d}$
\begin{equation}
\rho(r,z) = \frac{M_\mathrm{d}}{4\pi R_\mathrm{d}^2 z_\mathrm{d}} \cdot \exp(-R/R_\mathrm{d}) \cdot \sech^2(z/z_\mathrm{d}) \,.
\label{eq:sigma_disk} 
\end{equation}
The dark halo has a NFW-like profile \citep{NFW}:
\begin{equation}
\rho = 
\frac{C_\mathrm{h}\,T(r/r_\mathrm{t})}
{(r/r_\mathrm{s})^{\gamma_0}
\left((r/r_\mathrm{s})^{\eta}+1\right)^
{(\gamma_{\infty}-\gamma_0)/\eta}} \,,
\label{eq:NFW}
\end{equation}
where $r_\mathrm{s}$ is the halo scale radius, $r_\mathrm{t}$ is the halo truncation radius, $\eta$ is the halo transition exponent, $\gamma_0$ is the halo inner logarithmic density slope, $\gamma_{\infty}$ is the halo outer logarithmic density slope, $C_\mathrm{h}$ is the parameter defining the full mass of the halo $M_\mathrm{h}$. $T(x)$ is the truncation function with $x=r/r_\mathrm{t}$: 
\begin{equation}
T(x) = \frac{2}{\sech{x} + 1/\sech{x}} \,.
\end{equation}
We use $\eta=4/9$, $\gamma_0=7/9$, $\gamma_{\infty}=31/9$, which makes the halo density profile very similar to the NFW profile with a slightly steeper slope in the inner region.
The parameters were chosen so that the mass of the dark halo $M_\mathrm{h}(R<4R_\mathrm{d})$ within $4 R_\mathrm{d}$ was $1.5 M_\mathrm{d}$.
Each component is represented by a set of particles, $N_\mathrm{d}=4 \cdot 10^6$ for the disk and $N_\mathrm{h}=4.5 \cdot 10^6$ for the halo. The disk and the halo are self-consistent, i.e. all components evolve under the influence of their mutual gravitational field.
\par 
Disk parameters are fixed by the choice of the system of units: 
$M_\mathrm{d}=1, \, R_\mathrm{d}=1$. The disk thickness $z_{\mathrm{d}}=0.05$. For simplicity, we also set $G=1$. If one uses $R_\mathrm{d}=3.5$ kpc and $M_\mathrm{d}=10^{10} M_{\sun}$ then the time unit will be 
$t_\mathrm{u}=13.2$ Myr. 
\par 
The initial velocity dispersion profile $\sigma_R (R)$ of the disk was exponential with a scale $R_\sigma$ equal to $2 R_\mathrm{d}$
\begin{equation}
\sigma_R = \sigma_0 \cdot \exp(-R/2 R_\mathrm{d}).
\end{equation}
The constant $\sigma_0$ is determined in such a way that the value of the Toomre parameter at $R_\sigma$ $Q_{\mathrm{T}}(R_\sigma)$ is equal to some initially chosen $Q$. For our two models, we use $Q=1.2$ and $Q=1.6$, respectively. During its secular evolution the second model produces double X-structures \citep{Smirnov_Sotnikova2018} and we denote the first model as `SX' (single) and the second one as `DX' (double).
\par
The initial equilibrium state was prepared via a script for constructing the equilibrium multicomponent model of a galaxy {\tt{mkgalaxy}} \citep{McMillan_Dehnen2007} from the toolbox for N-body simulation {\tt{NEMO}} \citep{Teuben_1995}. 

\subsection{Simulations}
The evolution of modelled galaxies was followed using the fastest $N$-body code for one CPU {\tt{gyrfalcON}} \citep{Dehnen2002}, which combines a hierarchical tree method \citep{Barnes_Hut1986} and a fast multipole method \citep{Greengard_Rokhlin1987}. 
This code has complexity ${\cal O}(N)$. Its implementation was also taken from the public {\tt{NEMO}} package \citep{Teuben_1995}.
The adopted softening lengths were, respectively, 0.004 and 0.013 for the disk and the halo. We used the tolerance parameter $\theta=0.6$.
\par
\begin{figure*}
\begin{minipage}[t]{1.0 \textwidth}
\includegraphics[scale=0.37]{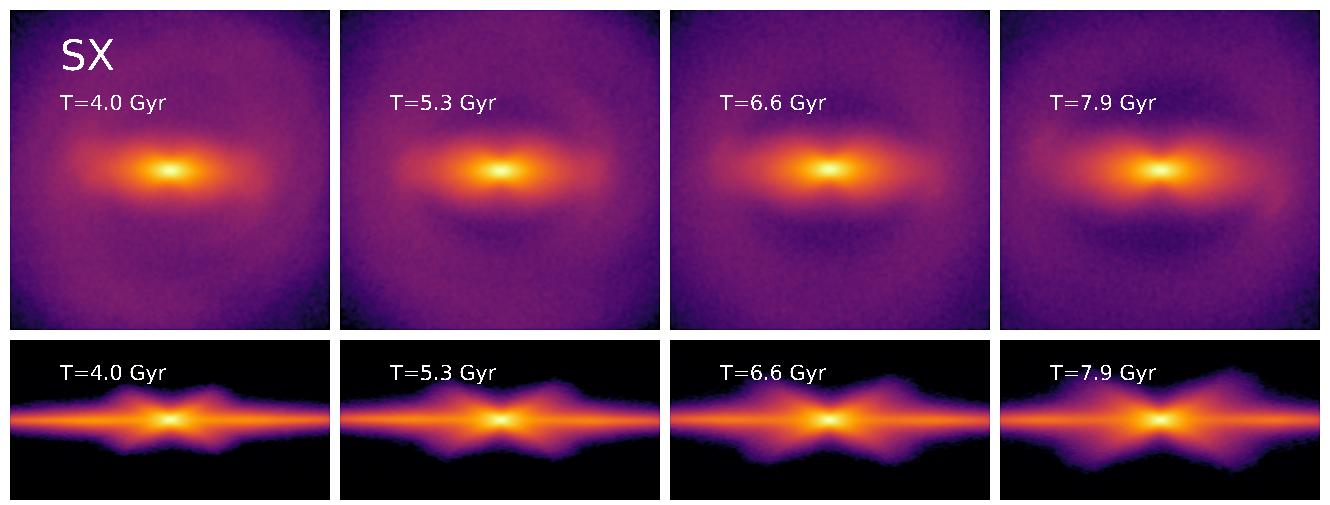}
\end{minipage}
\begin{minipage}[t]{1.0 \textwidth}
\includegraphics[scale=0.37]{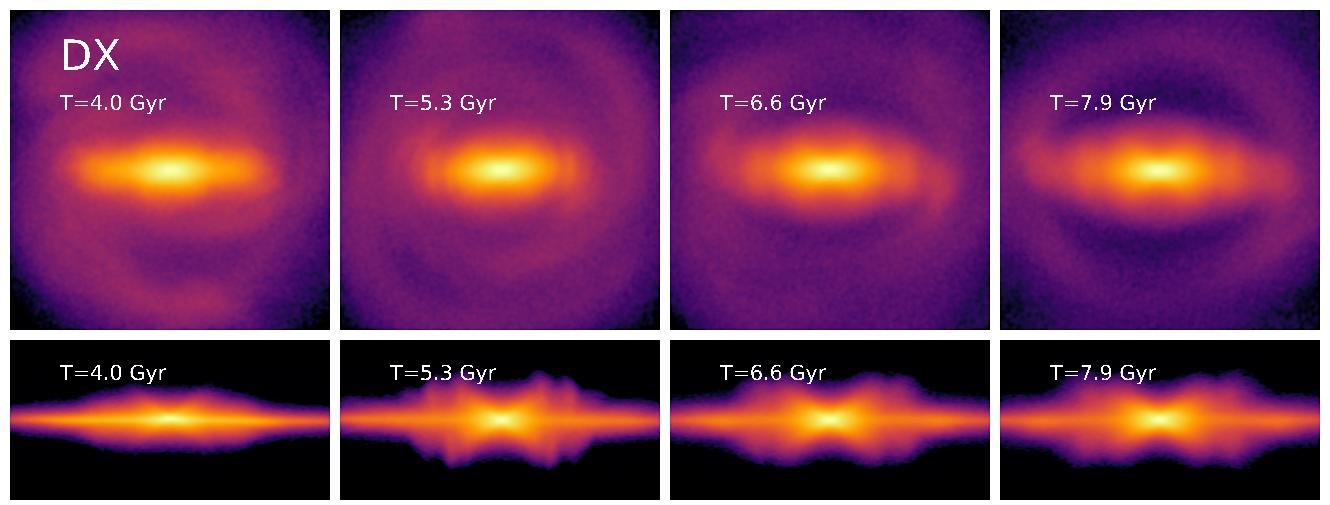}
\end{minipage}
\caption{The \textcolor{black}{face-on} and edge-on views of SX (two first rows from the top) and DX (third and fourth rows) models with \textcolor{black}{a bar seen side-on at four different time moments $t=300\,(4.0\, \mathrm{Gyr}),\,400\,(5.3\, \mathrm{Gyr}),\,500\,(6.6\, \mathrm{Gyr}),\,600\,(7.9\, \mathrm{Gyr})$. Face-on view is displayed in the square $(xy) = (-4,4) \times (-4,4)$ while edge-on view is displayed in the rectangle $(xz) = (-4,4)\times(-2,2)$.}
}
\label{fig:XY_XZ}
\end{figure*}

We re-ran the simulations from $t=300$ ($\approx 4$~Gyr) up to $t=600$ ($\approx 8$~Gyr) and saved outputs every 0.125 time units ($\approx 1.65$~Myr), i.e., forty times more frequently than in \citet{Smirnov_Sotnikova2018}. The general evolution of both models during this time period is shown in Fig.~\ref{fig:XY_XZ}. By $t=300$ the bar in both models is well developed both in and across the disk plane. At this time moment, the B/PS bulges and its X-shaped substructures are also clearly distinguishable. The model SX with $Q=1.2$ is already vertically symmetric, while the model with $Q=1.6$ experiences a strong secondary buckling at $t=350$ but by $t=400$ becomes vertically symmetric (see figure~17 in \citealt{Smirnov_Sotnikova2018}). The buckling itself has a striking appearance if the bar seen side-on (figure~10 in \citealt{Smirnov_Sotnikova2018}). \textcolor{black}{During the buckling phase four additional large bright rays rise from the bar plane. The rays are strongly asymmetric with respect to the disk plane. After the buckling, these rays are smeared and form the density enhancements slightly farther from the disk center after the first X-structure (see $t=6.6$ Gyrs for DX model in Fig.~\ref{fig:XY_XZ}). 
These density enhancements seem to be connected with the bar. They look like four lobes, which are similar to the first X-structure but are more blurry. Since they are remnants of four large and bright rays from the buckling phase we call them  ``the secondary X-structure''. Such structures are not a feature of our models only. \citet{Debattista_etal2017} also gives an example of double X-structures in their in-plane hot enough model (see their figure 8).}  
\par
\begin{figure}
\includegraphics[scale=0.52]{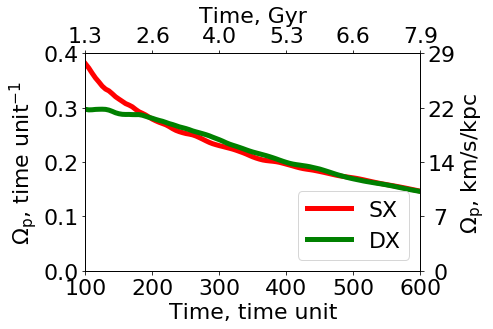}
\caption{Evolution of the bar pattern speed for SX and DX models.}
\label{fig:omega_p}
\end{figure}

For further analysis of \textcolor{black}{the} orbital structure it is crucial to understand how the bar pattern speed varies with time. We derive the pattern speed from the evolution of the bar position angle, \mbox{$\Omega_\mathrm{p}= d\psi/(2\cdot dt)$}, where the position angle is calculated via Fourier transform (FT) of the density of the disk,
\mbox{$\psi=\arctan\left(\Im(A_2)/\Re(A_2)\right)$}. The evolution of the pattern speed in \textcolor{black}{inverse time units of our simulations (``t.u.'' for short) and in commonly used pattern speed units, km/s/kpc,} is shown in Fig.~\ref{fig:omega_p}. \textcolor{black}{One inverse t.u. is approximately equal to $72$ km/s/kpc. As can be seen in Fig.~\ref{fig:omega_p}}, at the time interval $t=100-600$ the bar in both models is gradually slowing down. \textcolor{black}{Simultaneously  it is growing in size in both directions, radial and vertical (Fig.~\ref{fig:XY_XZ}).} 

\section{Analysis of dominant frequencies}

\begin{figure*}
\begin{minipage}[t]{0.5 \textwidth}
\includegraphics[scale=0.5]{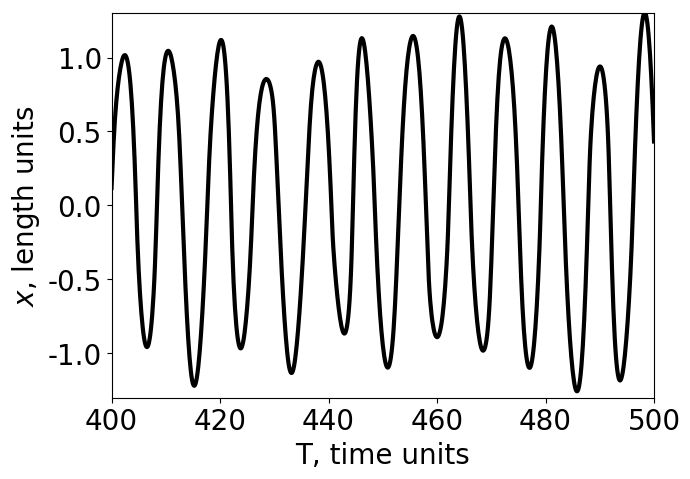}%
\end{minipage}%
\begin{minipage}[t]{0.5 \textwidth}
\includegraphics[scale=0.5]{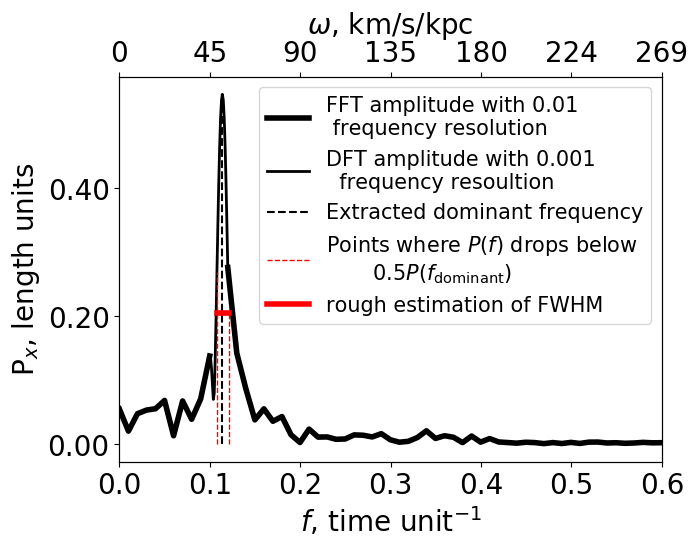}%
\end{minipage}%
\begin{minipage}[t]{0.24\textwidth}
\includegraphics[scale=0.25]{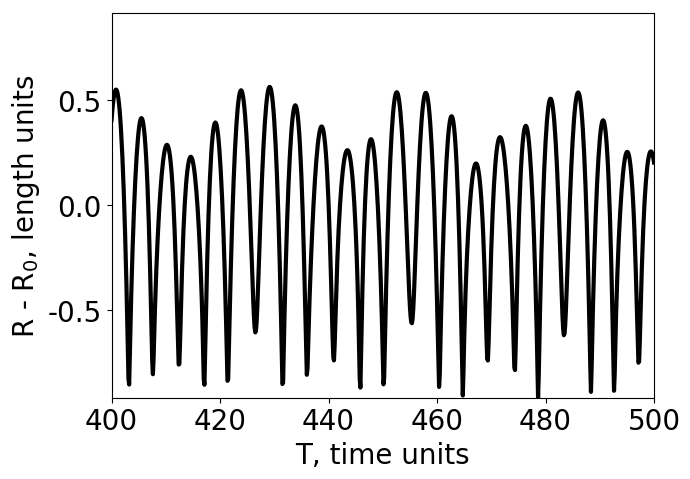}%
\end{minipage}%
\begin{minipage}[t]{0.24 \textwidth}
\includegraphics[scale=0.25]{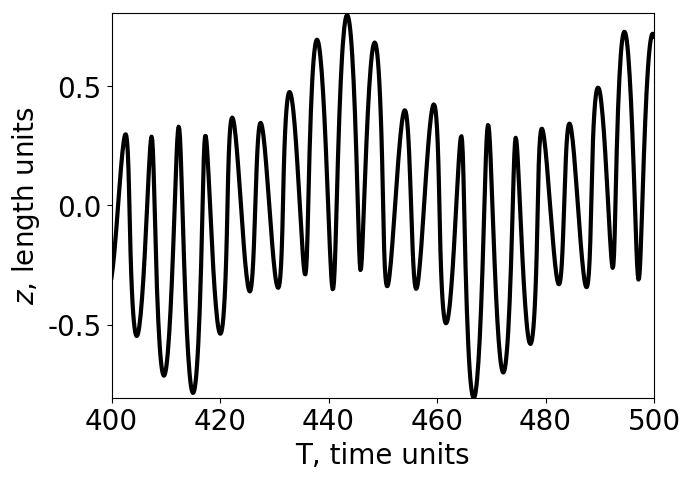}%
\end{minipage}
\begin{minipage}[t]{0.24\textwidth}
\includegraphics[scale=0.25]{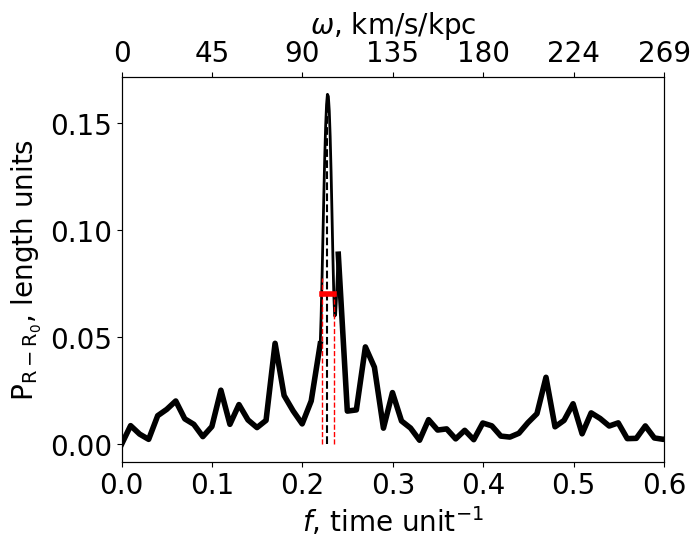}%
\end{minipage}%
\begin{minipage}[t]{0.24 \textwidth}
\includegraphics[scale=0.25]{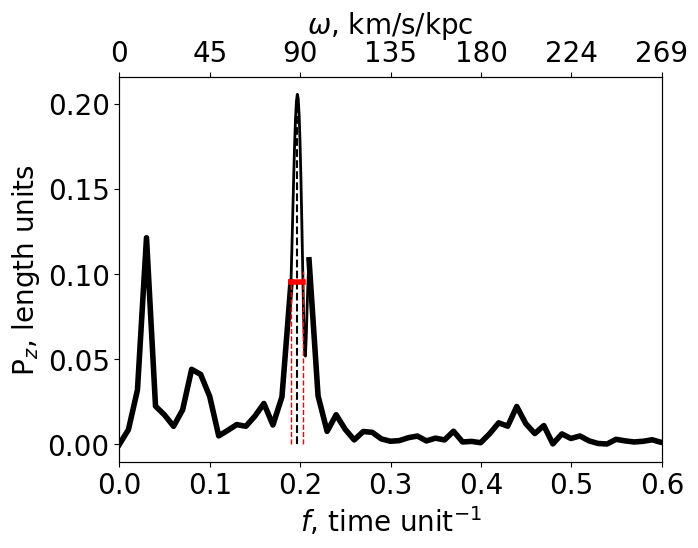}%
\end{minipage}%
\caption{Example of measurement of three different frequencies $f_x$, $f_\mathrm{R}$ and $f_z$ used to study orbital families. \textit{Left}: variation with time of corresponding coordinates. \textit{Right}: Periodogram of each time series from the left panel respectively, with various details of the present analysis.}
\label{fig:fourier}
\end{figure*}

\begin{figure}
\begin{minipage}[t]{0.2 \textwidth}
\raisebox{-\height}{\includegraphics[scale=0.18]{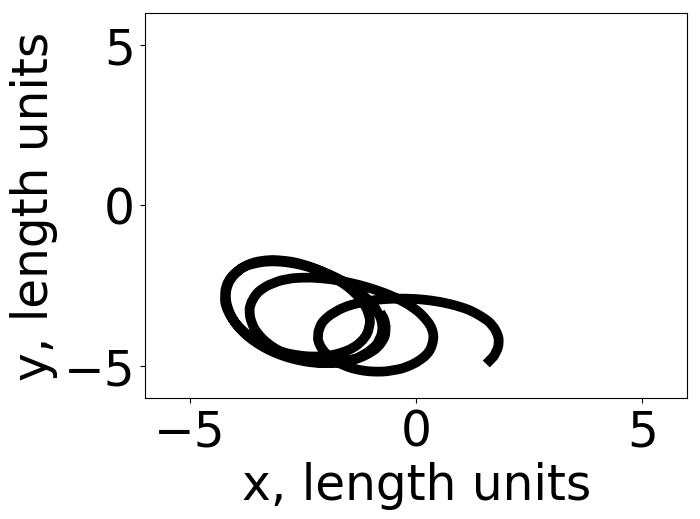}}
\raisebox{-\height}{\includegraphics[scale=0.18]{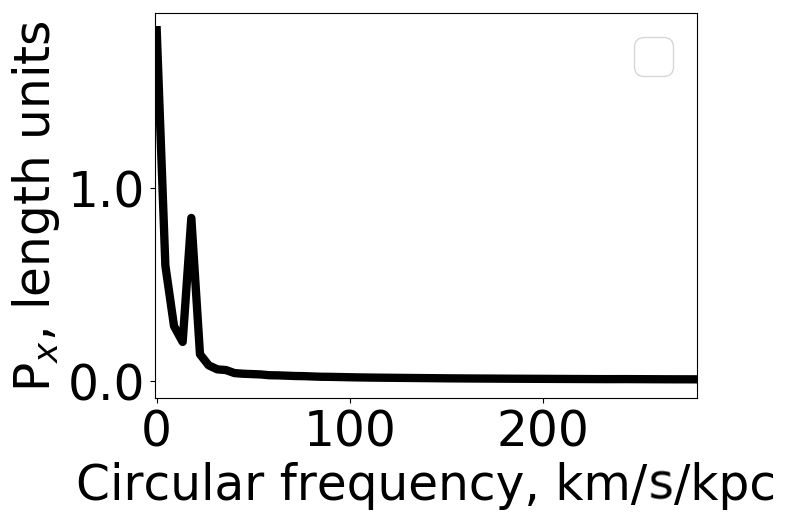}}%
\end{minipage}
\hfill
\begin{minipage}[t]{0.5 \textwidth}
\raisebox{-\height}{\includegraphics[scale=0.35]{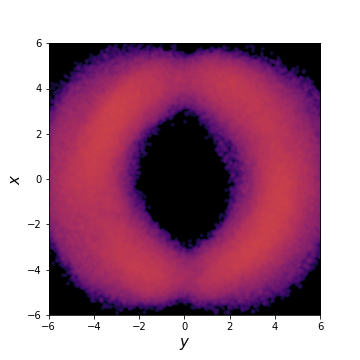}}
\end{minipage}
\caption{Particles with the dominant frequency $f_x=0$. \textit{Left column}: an example of a typical orbit in the plane XY (\textit{top}) and the corresponding periodogram $P_x$ (\textit{bottom}). \textit{Right column}: the snapshot of all such particles in SX model at $t=450$. We note, that the color normalization is the same as in Fig.~\ref{fig:XY_XZ}.
These particles were cut out from the present analysis.}
\label{fig:fx0}
\end{figure}

\begin{figure}
\center{\includegraphics[scale=0.45]{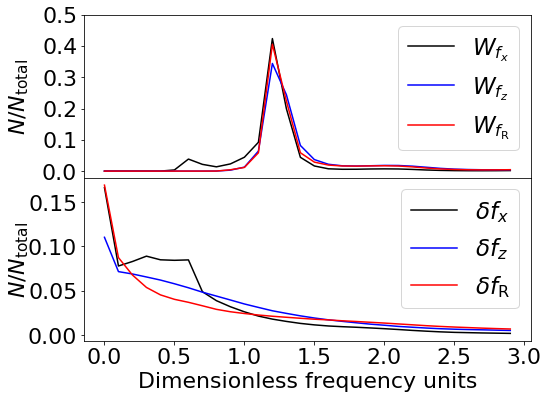}}
\caption{Distribution of particles over the dimensionless line width $W_f \cdot \Delta T$ (\textit{top}) and frequency shifts $\delta f \cdot \Delta T$ (\textit{bottom}) for different frequencies, where $\Delta T$ is time length of our time series. The bin width is equal to $\Delta f \Delta T=0.1$.}
\label{fig:wf}
\end{figure}



We consider the orbital motion of particles in a reference frame co-rotating with a bar. In this reference frame, $x$-axis goes along the major axis of the bar, which rotates with the pattern speed  $\Omega_\mathrm{p}$, and $z$-axis is orthogonal to the galactic plane ($y$-axis is determined by the right-hand rule). \textcolor{black}{Such a reference frame is a natural choice for a rotating bar because its regular backbone orbits from x1 family become closed in a disk plane.} 
\textcolor{black}{Further we use the cylindrical radius $R=\sqrt{x^2+y^2}$.}
\par 
To classify the orbits that support X-structures, we characterize each orbit in terms of frequency ratios $f_\mathrm{R}/f_x$ and $f_z/f_x$, where each frequency is the dominant frequency (the frequency of the line with the highest amplitude) in the corresponding coordinate spectrum. The frequencies were obtained as follows. We choose the time interval $t=400-500$ \textcolor{black}{where the pattern speed of the bar evolves slowly. The fractional change of the pattern speed at this time interval is approximately equal to $|\Omega_\mathrm{p}(t=500)-\Omega_\mathrm{p}(t=400)|/\Omega_\mathrm{p}(t=450) \approx 20\% $ (see Fig.~\ref{fig:omega_p})}. We assume that there is no significant change in orbital frequencies as the bar evolves slowly.   
\textcolor{black}{Below we calculate the frequency values at adjacent time intervals and show that our assumption is correct (see the end of this section). For the time interval $t=400-500$,} the coordinates of all disk particles with the time step $\Delta t=0.125$ were taken, so that for each particle a time series of $N_\mathrm{t}=801$ positions in the Cartesian coordinate system was formed. The dominant frequency was identified as the frequency corresponding to the highest peak of the periodogram obtained as a result of applying the fast Fourier transform (FFT) algorithm. 
\par
The periodogram $P_x$ for coordinate $x$ is calculated as follows
\begin{equation}
P_x =  \frac{1}{N_t} \left|\sum_{k=0}^{k=N_t-1} x_k  
\exp (-2\pi i f_{j} t_k) \right|, 
\label{eq:FT}
\end{equation}
where  $f_{j} = j/\Delta T$, $\Delta T = 100.125$, $t_k = k \Delta t$ and $0\leq j \leq (N_t-1)/2$. Periodograms $P_\mathrm{R}$ and $P_z$ are calculated in the same way. 
\par
Fig.~\ref{fig:fourier} shows the changes in coordinates over time, $x(t)$, $R(t)$ and $z(t)$, and the corresponding periodograms for one of the typical bar particles. The figure also shows various important details of the present analysis which will be discussed below.
\par
For almost all particles, the dominant frequency $f_\mathrm{R}$ is equal to zero since stars usually rotate around the center and do not fall directly into it. For some particles, the dominant frequencies $f_x$ and $f_z$ are also equal to zero. In the case of $f_x$, these particles are disk particles that have not enough time to revolve once around the center (see Fig.~\ref{fig:fx0}). Consequently, they do not belong to the bar and we removed such particles from further analysis. Each model has about $4 \cdot 10^5$ of such particles. In the case of $f_z$, a visual inspection of orbits showed that they are banana-shaped orbits and, therefore, belong to the bar. To account for particles with $f_\mathrm{R}=0$ and $f_z=0$, we subtracted the mean values of $<z>_t=z_0$ and $<\mathrm{R}>_t=\mathrm{R}_0$ for all particles at the pre-processing stage and further used the dominant frequencies of periodograms obtained from this modified time series. 
\par
Due to the discreteness of the time series under study, the maximal value of the frequency that can be determined is limited by the Nyquist frequency $\displaystyle \nu_\mathrm{c} = \frac{1}{2\,\Delta t} = 4.0$ t.u.$^{-1}$, or $\omega_c = 2\pi \nu_c \approx 1800$ km/s/kpc. The limited length of the time series leads to a loss of resolution, which in this case is calculated using the formula $\displaystyle \Delta \nu = \frac{1}{N_\mathrm{t}\, \Delta t} = 0.01$\footnote{The exact value of frequency step $\Delta \nu$ is slightly smaller because $N_t$ is an odd number, $\Delta \nu =1.0/100.125\approx 0.0099875$ t.u.$^{-1}$.} t.u.$^{-1}$, or $\Delta \omega \approx 4.5$ km/s/kpc. In FFT, this frequency also serves as the frequency resolution of the peak frequency. In our case, we found this resolution insufficient for accurate measurement of resonance ratios. In general, the peak resolution can be improved using various methods, \textcolor{black}{for example, by assuming some peak shape (Gaussian or parabolic, see \citealp{Gasior_Gonzalez04}) or by applying FT to zero-padded time series (see \citealp{Lyons}, chapter 3.11). In short, FT of zero-padded time series provides a better resolution of the peak frequency as the resulting frequency grid is denser compared to that of the usual FFT. An important feature of the outcome of this procedure is that it increases the resolution of the peak frequency if there is only one peak, but it does not provide a better frequency resolution for distinguishing two nearby lines. The later can be achieved only by increasing the length of the time series. Therefore, the initial time series should be large enough to distinguish between different lines in the orbit spectrum. In particular, the Nyquist frequency $\nu_c$ should be above the highest observed frequency in order to avoid the overlapping of different spectral lines (see \citealp{Lyons}, chapter 2.1). The small time step of our time series ensures this. As we show in the following sections, the typical frequencies of orbits are limited by about 1.0 t.u.$^{-1}$ which is well below the Nyquist frequency $\nu_c=4.0$ t.u$^{-1}$ for our time series.
In the present analysis, we use the procedure similar to zero-padding one but in a modified version as we are mainly interested in only one frequency peak. The details of our algorithm are as follows.} 
\par
First, we found a rough approximation of the peak frequency $f_j$ which corresponds to the highest line from the usual FFT periodogram. Then we formed a ten times more dense grid with frequency step $\Delta f = 0.001$ t.u.$^{-1}$ around this peak from one previous $f_{j-1}$ to the next node $f_{j+1}$ of the base FFT frequencies. For each node of this newly formed grid, we calculated discrete Fourier transform (DFT) of our time series (\textcolor{black}{thin black line in the top right plot} of Fig.~\ref{fig:fourier}). The maximum of this new DFT periodogram is a more accurate \textcolor{black}{estimate} of the dominant frequency, which we used in further analysis. \textcolor{black}{The final frequency resolution of the peak which we obtain after applying the above procedure is $\Delta f=0.001$ t.u.$^{-1}$. We note, that \citet{Gajda_2016} also calculated spectra for a discrete set of frequencies with the spacing, which was four times denser than in the case of the classic FFT algorithm. They justify the use of a denser sampling by the fact that the FFT sampling is of the order of the width of an individual peak. Our algorithm uses the denser spacing only in the vicinity of peaks.}
\par 
To ensure that even with such a small resolution, we trace each line with at least a few points, we roughly estimated the width of the lines. This was done in the following way. Each dominant frequency $f_k$ has a corresponding line amplitude $A_k$. Starting from the frequency $f_k$, we descend from the peak of the line to lower frequencies with a step $\Delta f$ and calculate the corresponding amplitude $P(f)$ using Eq.~\eqref{eq:FT}. After some iterations, we find the frequency $f_1 < f_j$ such that $P(f_1) \leq A_k/2$ and $P(f_1+\Delta f)> A_k/2$. Similarly, we find the frequency $f_2$ such that $P(f_2) < A_k/2$, $P(f_2 - \Delta f) > A_k/2$ and $f_2 > f_j$. Thus, the difference $W_f = f_2 - f_1 $ can be taken as a rough estimation of the full width at half maximum (FWHM). The resulting $f_1$, $f_2$ and $W_f$ for periodograms of our probe particle are also shown in Fig.~\ref{fig:fourier} (dashed and thick red lines, respectively).
\par 
Fig.~\ref{fig:wf} shows the distribution of all particles from SX model over the quantity $W_f \cdot \Delta T$, which is the line width in units of usual FFT resolution frequency $\Delta \nu$. One can see that the typical value of this quantity is close to unity. In other words, the typical width of the lines is about $1/\Delta T$. Consequently, we trace each line with at least ten points in our modified algorithm. We note, that this picture also shows that the usual FFT resolution turns out to be insufficient for determining the peak frequency, as we mentioned earlier.
\par 
\textcolor{black}{To verify our initial assumption of small frequency shifts at time interval $400-500$ we calculated the peak frequencies at adjacent time intervals $t=350-450$ and $t=450-550$. Thus the frequency shift of individual orbits can be estimated as the difference between peak frequencies at these time intervals, $\delta f_i = |f_{i,450-550} - f_{i,350-450}|$, where $i$ stands for $x,y,z$ or $\mathrm{R}$. The distribution of orbits over the frequency shifts $f_i$ is shown in Fig.~\ref{fig:wf}. A typical frequency shift for our orbits is less than $0.005$ t.u.$^{-1}$, although there are small number of orbits with $\delta f > 0.200$ t.u.$^{-1}$. From Fig.~\ref{fig:wf}, we can conclude that for most particles our initial assumption of small frequency shifts is correct. It is also interesting that there are orbits with no detected frequency shifts even with our increased frequency resolution. Visual inspection of such orbits revealed that they come in a variety of types. Some of them are the orbits of bar particles from the most central region while others are circular orbits of the outermost disk particles. We also note that the frequency shifts obtained by FT of the adjacent time intervals are the upper limits for actual frequency shifts from our main time interval, $t=400-500$. The reason for this is the larger time derivative of the pattern speed at the time interval $t=350-450$ compared to that at the interval $t=400-500$ (see Fig.~\ref{fig:omega_p}). } 
\par
Subsequently, \textcolor{black}{various types of regular} orbits were distinguished based on the analysis of frequency ratios $f_\mathrm{R}/f_x$ and $f_z/f_x$. \textcolor{black}{Since we estimate frequencies with some finite frequency resolution, there is no strict condition that allows us to distinguish between near-periodic and non-periodic but regular orbits on the closest invariant curves around ideal periodic ones. It is a natural drawback of spectral analysis of the orbits arising in $N$-body simulations and thus cannot be avoided. We also note that such an analysis does not allow us to distinguish between sticky chaotic and regular orbits. \cite{Wang_etal2016} showed that the fraction of orbits, which can be claimed chaotic, is strongly dependent on some predefined parameters such as the maximum allowed frequency drift for regular orbits and the maximal number of lines which we are trying to identify in their spectra. However, \cite{Machado_Manos2016}, using a different approach, showed that the fraction of chaotic orbits in the bar decreases with time, reaching about 10\% after 6 Gyrs of evolution for the models with a strong bar. The bars in our models contain more than 50\% of all initial disc particles. Therefore, we assume that the contribution of chaotic orbits is small for our models with a strong bar and a non-strict accounting \textcolor{black}{for} them does not affect the results of our analysis.} 
\section{Bar and disk} 
\label{sec:bar_disk}

\begin{figure*}
\centering
\begin{minipage}{.48\textwidth}
\centering
\includegraphics[width=\textwidth]{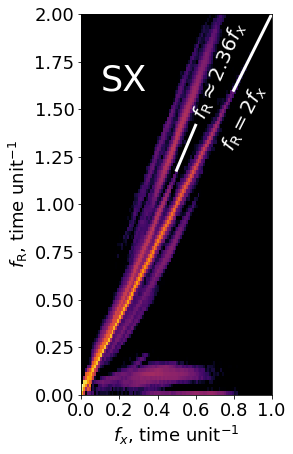}\\
\end{minipage}
\hfill
\begin{minipage}{.48\textwidth}
\centering
\includegraphics[width=\textwidth]{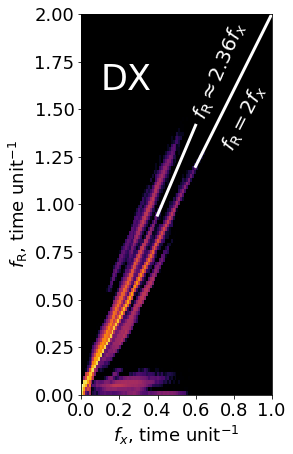}\\
\end{minipage}
\caption{2D distribution of particles over $f_\mathrm{R}$ and $f_x$ frequencies for SX (\textit{left}) and DX (\textit{right}) models. \textcolor{black}{White lines mark the most populated orbital families: an orbital family with the bar frequency ratio $f_\mathrm{R}/f_x = 2$ and an additional orbital family with $f_\mathrm{R}/f_x \approx 2.36$ ($f_x/f_\mathrm{R} \approx 0.42)$.}}
\label{fig:fRx}
\end{figure*}
Since the X-structure is a natural part of the B/PS bulge, we begin by analyzing the orbital composition of the whole B/PS bulge and then proceed with the analysis of the X-structure itself. In the first step, we separate the \textcolor{black}{3D bar} particles from the disk particles. The orbits of particles in the disk are close to circular, and these particles do not contribute to the particular vertical structure of the models. 
\textcolor{black}{On the opposite, the B/PS bulge is mainly populated by different kinds of regular/sticky chaotic 3D orbits. These orbits arise due to various dynamic mechanisms but most of them are connected with x1 orbits family tree (see \citealt{Patsis_Katsanikas2014a} for a detailed review). The x1 family itself is elongated along the 2D bar and \textcolor{black}{constitutes} its backbone. As the bar grows, this family becomes larger and larger. At the same time, some of its new members deviate from a mid-plane and become a new part of a peanut structure, thereby further building \textcolor{black}{it} up \citep{Gajda_2016,Lokas2019}.}
\par 
The characteristic that allows to separate disk particles from the particles of the B/PS bulge is the frequency ratio $f_\mathrm{R}/f_x$.
\textcolor{black}{But before analyzing the various groups of orbits based on their frequency ratio, we would like to make a few remarks regarding the terminology used below.
There are different classifications of orbits in the literature. In works where individual orbits are studied in terms of Poincare surface sections, the term family is introduced. The family refers to the set of periodic orbits which can be found by continuously varying some physical parameters of the orbits (for example, the Jacobi energy of the orbit, (see \citealt{Contopoulos2002}, chapter 2.4.2 for a review)). Orbits from the same family are characterized by the same number of intersection points with Poincare sections. In works that deal with numerical simulations, there is a tendency toward a simpler classification based on the general morphology of an orbit (see \citealt{Valluri_etal2016}, for example). In the present work, we adopt another classification of orbits based on the spectra of the orbits and their properties. In particular, we use terms ``group'' or ``family'' of the orbits to refer to the set of orbits with nearly the same frequency ratios $f_\mathrm{R}/f_x$ or $f_z/f_x$. Our approach is close to that used in the work by \citet{Portail_etal2015a}. In practice, ``nearly the same'' means that the frequency ratios within one group of orbits differ by less than $0.1$. Such a classification is somewhat arbitrary, since real orbit distributions continuously span a certain range of frequency ratios (see Sec.~\ref{sec:v_res}). However, such division allows us to roughly evaluate the role of various types of orbits in the assembly of the X-structure arising in such a self-consistent potential as ours. Where possible, we try to compare our results with previous works, but such a comparison should be interpreted with great caution since the classification methodology across different works is different.}
\par 
 Fig.~\ref{fig:fRx} shows the 2D distribution of the frequencies $f_x$ and $f_\mathrm{R}$ for SX and DX models. 
The family of particles with $f_\mathrm{R}/f_x$ in the range $f_\mathrm{R}/f_x \in 2.0 \pm \Delta (f_\mathrm{R}/f_x)$ can be clearly distinguished (the brightest strip in both pictures). These particles constitute the backbone of the B/PS bulge, 
which is essentially a 3D bar that has grown in the vertical direction. 
\textcolor{black}{About 50\% of all particles that represented the initial disk are involved in it in both models.} \textcolor{black}{The visual inspection of orbits showed that the condition imposed on $f_\mathrm{R}/f_x$ does not allow one to distinguish between  $x1$ orbits and box orbits \citep{Valluri_etal2016} which also populate the bar. To do this, one needs to classify orbits based on the ratio of the frequencies $f_y/f_x$ which \textcolor{black}{characterizes} the oscillations of a star in the disc plane. In the future, we plan to analyze the `face-on portrait' of the bar, where we will conduct a more \textcolor{black}{detailed} classification of the orbits. In this work, we are mainly interested in the vertical bar structure and do not dig further \textcolor{black}{into} this question.}
\par 
A separate family of particles can be identified near $f_\mathrm{R}/f_x \simeq 2.4$.
Perhaps they can be associated with particles having 
$(\Omega - \Omega_\mathrm{p})/\kappa = 0.44$ revealed by \citet{Gajda_2016} (see also \citealp{Harsoula_Kalapotharakos2009}). \textcolor{black}{These particles do not belong} \textcolor{black}{to the outer disk}. \textcolor{black}{They partially contribute to the B/PS bulge but the orbits are not elongated along the bar. For these orbits, the frequency ratio $f_y/f_x$ is grouped around a value of 1.33. A few of them are near periodic and look very similar\footnote{Perhaps they are related to the orbits in figure~A6 in \citet{Patsis_Athanassoula2019}.} to the orbit in figure~6, row (c) in \citet{Gajda_2016} or to its `pretzel'-like version. Both types are found not only in the most central areas of the bar but also at its periphery. Most of the orbits from this family are unwind versions of these two types and \textcolor{black}{when} combined together form a roundish shape in a face-on view.} This family is practically absent in the model SX ($\approx$ 5\%), although in DX model such particles are about 20\%.
The remaining families (for example, second strip from the top of the picture) are even smaller in number. On average, their orbits have large $f_x$ and contribute only in the central areas of a galaxy. The outer disk (outside the bar) is inhabited by orbits with very low $f_\mathrm{R}$ and $f_x$ (a bright spot near zero).
As for the bar particles, \citet{Portail_etal2015b} identified them by their frequency ratio as $f_\mathrm{R}/f_x \in 2.0 \pm 0.1$.
For further analysis, we used the same condition to isolate the particles involved in the bar.

\section{Orbital components of the peanut structure} 
\label{sec:v_res}

\begin{figure*}
\centering
\begin{minipage}{.48\textwidth}
\centering
\includegraphics[width=\textwidth]{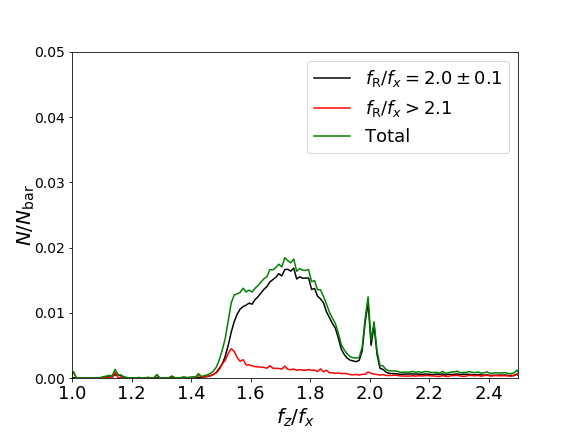}
\end{minipage}
\hfill
\begin{minipage}{.48\textwidth}
\centering
\includegraphics[width=\textwidth]{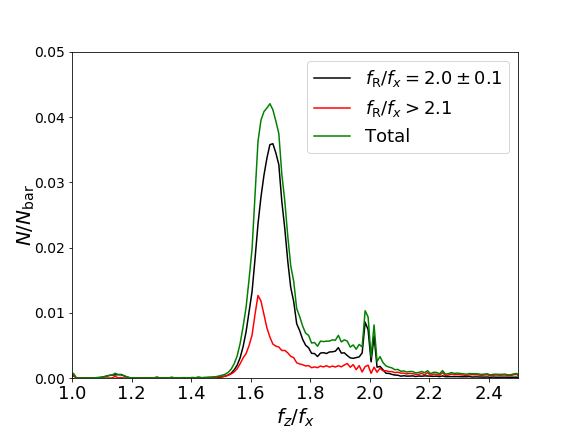}
\end{minipage}
\caption{The distribution of bar particles over the ratio $f_z/f_x$ for SX (\textit{left}) and DX models (\textit{right}) calculated for the time interval $t=400-500$. Black line --- all \textcolor{black}{bar} particles; red line --- particles with the ratio of epicyclic to $x$ oscillations frequencies $f_\mathrm{R}/f_x > 2.1$; green line --- the ratio $f_z/f_x$ calculated for both families of orbits.}
\label{fig:fzx}
\end{figure*}

\begin{figure}
\centering
\begin{minipage}{.49\textwidth}
\centering
\includegraphics[width=0.49\textwidth]{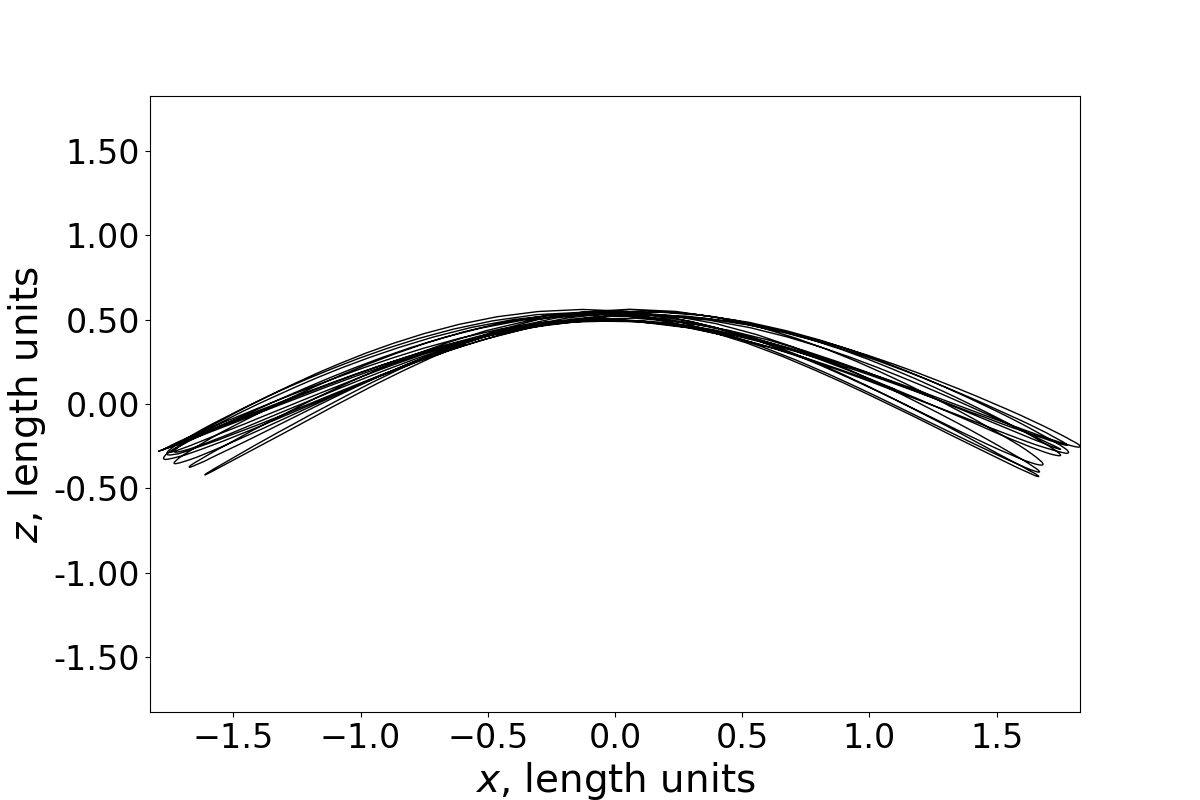}%
\centering
\includegraphics[width=0.49\textwidth]{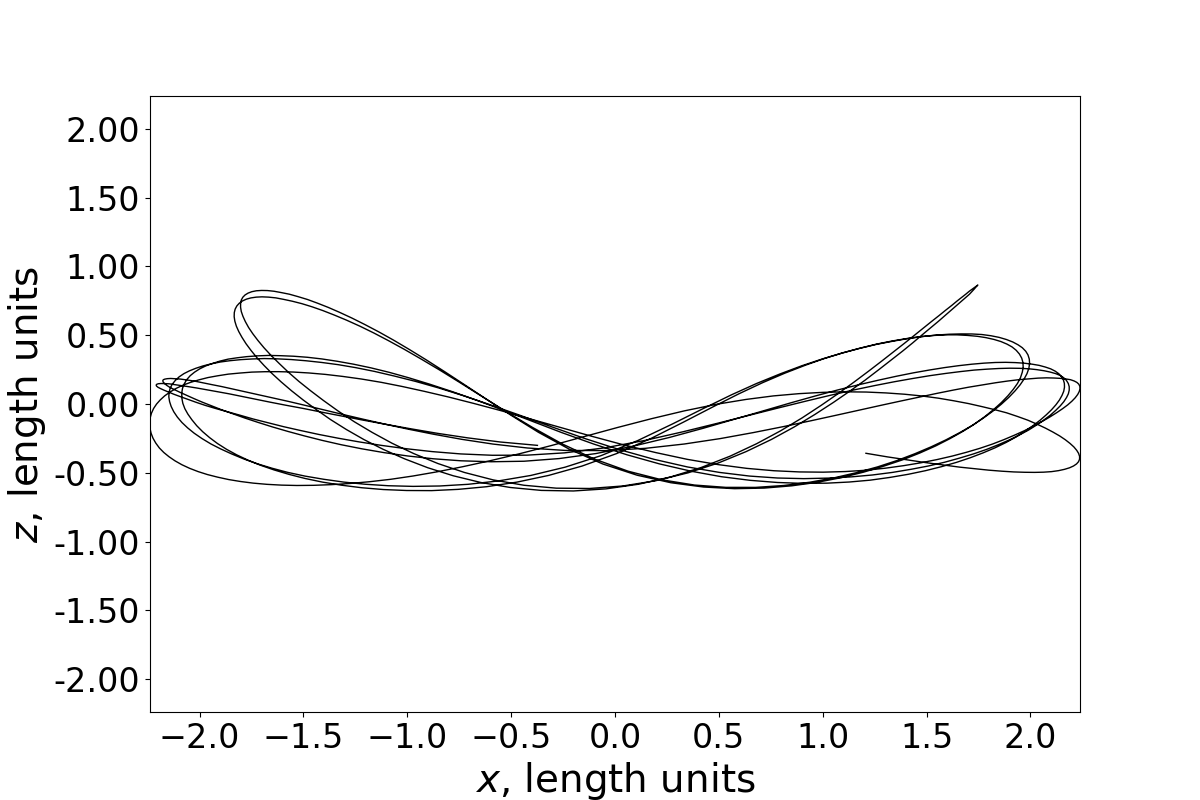}%
\end{minipage}
\caption{Orbits with $f_z/f_x \simeq 2.0$ for SX model.}
\label{fig:BAN_ABAN}
\end{figure}
The role of different families of orbits in the assembly of the vertical structure of the B/PS bar is determined by two points. The first one is the prevalence of each particular family of orbits with a certain ratio $f_z/f_x$. \textcolor{black}{It determines the contribution to the surface density made by particles from one family in comparison to the particles from other families.} The second one is the spatial distribution of such a family of orbits.
\textcolor{black}{For example, an ensemble of orbits can fill a certain area without creating noticeable density gradients, or, conversely, particles can be assembled in such a way that the overall density profile of the entire ensemble contains high-density gradient peculiarities like the rays of the X-structure.
For B/PS bulges, the former means that a certain family of orbits creates only the bulge background, without contributing to the X-structure. As we show below, this particular case is rare and all orbits from different families participate in the formation of the X-structure in our models.}
We begin this section with a general description of the distribution of the orbits over the ratio $f_z/f_x$. 
 In the following subsections, we describe the specific spatial distribution of different families of orbits and analyze which structure is assembled from each individual family.

\subsection{Distributions over the vertical to the in-plane frequencies ratio}
\textcolor{black}{Fig.~\ref{fig:fzx} shows the distribution over the ratio $f_z/f_x$ for bar particles with $1.9 \leq f_\mathrm{R}/f_x \leq 2.1$ (black line), for both models with a bin width equal to $\Delta (f_z/f_x)=0.01$.} The figure also shows the contribution of an additional family with the frequency ratio $f_\mathrm{R}/f_x \simeq 2.4$ (red line).
The orbits from this family are not elongated along the bar and contribute only to the very central regions of a B/PS bulge (see the previous section). Their contribution to the vertical structure in SX model is small and they affect only the low-frequency ratio part of the distribution. For DX model they do not distort the shape of the distribution, but only slightly raise the maximum. Given that this family is small in number and \textcolor{black}{its orbits differ from the typical bar particle orbits}, we decided to exclude such \textcolor{black}{orbits} from the analysis.
\par
The most notable features of the $f_z/f_x$ distribution of \textcolor{black}{bar particles} for both models are as follows.
\par 
First, as can be seen from Fig.~\ref{fig:fzx}, orbits located near the 2:1 resonance ratio constitute a small fraction \textcolor{black}{($\approx7\%$)} of all B/PS bulge orbits. Although they cause a visible peak at $f_z/f_x=2$, they do not prevail over other orbits in both models, especially in DX model. This fact points out that 2:1 resonant orbits cannot be the main ``ingredients for cooking'' the X-structure. To show it we analyze in detail the spatial distribution of different orbital families in the next subsection.
\par
\par 
\textcolor{black}{We also note that there are different sub-types of 2:1 orbits which populate the peak in Fig.~\ref{fig:fzx}. The 2:1 resonance is usually associated with periodic orbits from the x1v1 or x1v2 families \citep{Patsis_etal2002b}, or BAN/ABAN orbits \citep{Pfenniger_Friedli1991}. There are some studies where the role of quasi-periodic orbits/sticky chaotic orbits associated with these families \textcolor{black}{in the X-structure building} were investigated \citep{Patsis_Katsanikas2014a,Patsis_Harsoula2018}. We visually inspected several tens of 2:1 orbits from our SX model and found that although for some orbits the frequency ratio is 2.0 with good accuracy ($2.00\pm 0.05$), the orbits themselves do not look like a banana.
Most of the orbits have a second wave on the dependency $z(t)$, which manifests itself as a second peak in the corresponding periodogram (see Fig.~\ref{fig:fourier}, right column). The presence of such a wave leads to a change in the morphology of the banana orbit. The orbit on the XZ plane looks like it flaps its wings, then lifts them up, then drops down, and near the line $z=0$ it turns into a sign of infinity (ABAN orbit). Fig.~\ref{fig:BAN_ABAN} demonstrates two types of orbits with $f_z/f_x=2.0$ from our SX model: an almost periodic\footnote{This is rather q.p. orbit, i.e. a thick version of an x1v1 orbit.} `banana' (left panel) and a librating `banana' (right panel). This hybrid morphology of the librating `banana' closely resembles the morphology of sticky chaotic orbits studied by \citealt{Patsis_Katsanikas2014a} (see their figure~13).  An example of a  librating `banana' orbit can also be seen in \citet{Portail_etal2015b}, albeit for the frequency ratio $f_z/f_x=1.9$ (figure~2, class E in this work).}
\par
\textcolor{black}{The contribution of the 2:1 orbits is somewhat similar in both models. At the same time}, the other groups of orbits appear in quite a different manner in each model (see Fig.~\ref{fig:fzx}). The distribution over the frequency ratio in SX model has a wide hump from $f_z/f_x\simeq1.5$ to $f_z/f_x\simeq1.95$ with the maximum near 7:4. At the same time, DX model has a rather narrow and high peak in the region of small values of the frequency ratio (the maximum is approximately at 5:3). In general, orbits with a lower ratio of frequencies $f_z/f_x$ prevail in the hotter model DX, while in the model SX we observe a more uniform distribution in the frequencies ratio below 2:1.
Notably, the distributions for both models are smooth in the range from $1.5$ to $1.9$, and there are no sharp resonance peaks.

\subsection{Spatial distribution of different groups of orbits}

\begin{figure*}
\centering
\begin{minipage}{.48\textwidth}
\centering
\includegraphics[width=\textwidth]{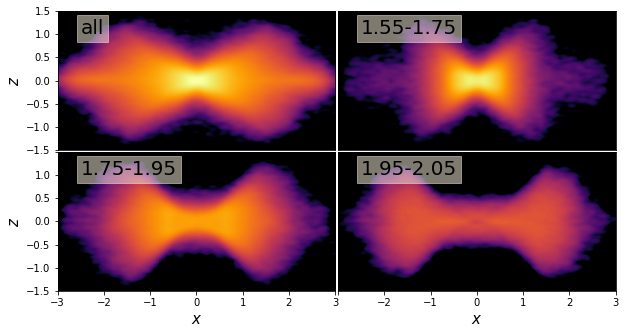}\\
\end{minipage}
\hfill
\begin{minipage}{.48\textwidth}
\centering
\includegraphics[width=\textwidth]{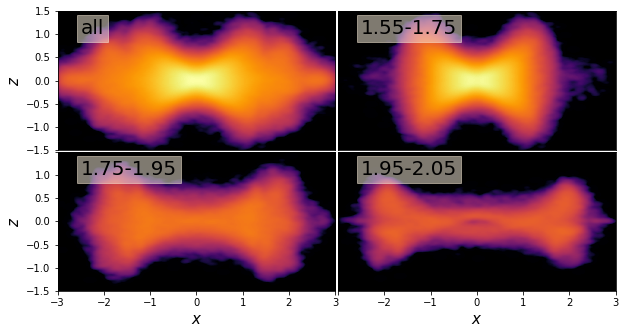}\\
\end{minipage}
\caption{The combined image of the B/PS bulge in the projection on the XZ plane and the contributions of individual groups of particles divided by the frequency ratio $f_z/f_x$ for the models SX (\textit{left}) and DX (\textit{right}). The edges of $f_z/f_x$ bins are signed at the top of small plots.}
\label{fig:xz_res}
\end{figure*}

Here we describe in detail what substructures of the B/PS bulge are assembled from different groups of orbits with different ratios of $f_z /f_x$.
\par
Fig.~\ref{fig:xz_res} presents snapshots for particles from three different parts of $f_z/f_x$ distribution for both models at $t=450$. The ``side-on'' view (the projection on the XZ plane) is shown, with the major axis of the bar lying in the picture plane. The division by the $f_z/f_x$ ratio was done as follows: $f_z/f_x = 1.55-1.75$, $f_z/f_x = 1.75-1.95$ and $f_z/f_x = 1.95-2.05$. The first images for the models SX and DX, respectively, are combined pictures consisting of all particles \textcolor{black}{with $f_\mathrm{R}/f_x = 2.0 \pm 0.1$}. The remaining images are ``portraits'' of different parts of $f_z/f_x$ distribution from low to higher values of the ratio $f_z/f_x$. 
\par 
As can be seen from the figures for each model, each considered part of the $f_z/f_x$ distribution exhibits a B/PS structure with a bright (for the low frequencies ratio part) and faint (for the high frequencies ratio part) X-shaped structures. The overall appearance of the selected areas differs significantly from each other and depends on their dominant orbits. Central areas look more like a `farfalle', or a `bow tie' pasta, while the outer regions have a clear bridge in the center and resemble a `dumbbell'. The length of the orbits with a low frequencies ratio, $f_z/f_x=1.55-1.75$, is approximately the same in the radial and vertical directions. 
Orbits with higher values of $f_z/f_x = 1.75-2.05$ are more elongated and fall closer to the disk plane. Banana orbits with $f_z/f_x=2.0$ contribute to the X-structure only in the outer regions and they just limit the length of the B/PS bulge in the radial direction. We note, that in our DX model, unlike SX model, there is a large number of 
`bananas' \textcolor{black}{like those depicted in Fig.~\ref{fig:BAN_ABAN} (left plot)}. They are well manifested in the form of two inverted crescents. Crescents, overlapping each other, form an empty space in the middle.
\par 
In general, Fig.~\ref{fig:xz_res} clearly demonstrates that the banana-shaped orbits cannot be the ``backbone'' of the X-structure \textcolor{black}{in our models.}. Moreover, other orbital groups seem to be tightly connected with the phenomenon of the X-structure.
\section{Contribution of different periodic and near-periodic orbits in the vertical structure}

\subsection{Jacobi integral}

A more detailed picture of the spatial distribution of orbits across the 3D bar can be obtained in the framework of the Jacobi integral, a constant that can be associated with each orbit in a non-inertial reference frame rotating at a constant angular speed. It is defined in the following way. Suppose that the probe bar particle with the absolute value of the velocity $v$ and the radius vector $\textbf{r}$ has the total energy $E=v^2/2+\Phi(\textbf{r})$, where $\Phi (\textbf{r})$ is the potential energy at the point $\textbf{r}$. Then the Jacobi constant \textcolor{black}{(energy)} is defined as $E_j = E - \Omega_\mathrm{p} \cdot L_z$, where $L_z$ is the vertical projection of the angular momentum and $\Omega_\mathrm{p}$ is the pattern speed of the bar. 
\par
\begin{figure}
\begin{minipage}[t]{0.48 \textwidth}
\includegraphics[width=\textwidth]{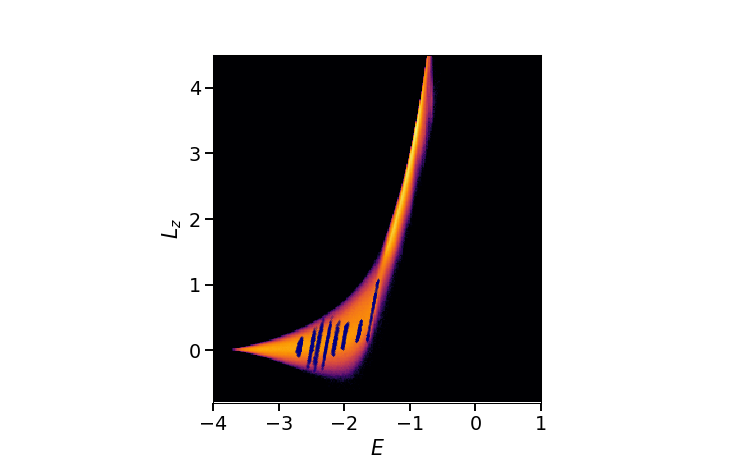}
\end{minipage}
\caption{Distribution of all particles of SX model on the plane total energy --- angular momentum $E - L_z$ and tracks of individual particles. The slope of tracks $k \approx 4.8$.}
\label{fig:ELz}
\end{figure}

Fig.~\ref{fig:ELz} shows the distribution of particles from the model SX at $t=500$ in coordinates $(E, \, L_z)$. The figure also contains the tracks of individual particles for the time interval $t=400-500$. It can be seen that during this time interval the particles move along almost straight lines with approximately the same slope, which varies very little with time. For both models, \textcolor{black}{each particle track gives a constant} 
$C_j=-E_j/ \Omega_\mathrm{p}=L_z - k E$, where the numerical coefficient $k=4.8$ follows from a specific value of the pattern speed for our models. 

\begin{figure*}
\begin{minipage}[t]{0.48 \textwidth}
\includegraphics[width=\textwidth]{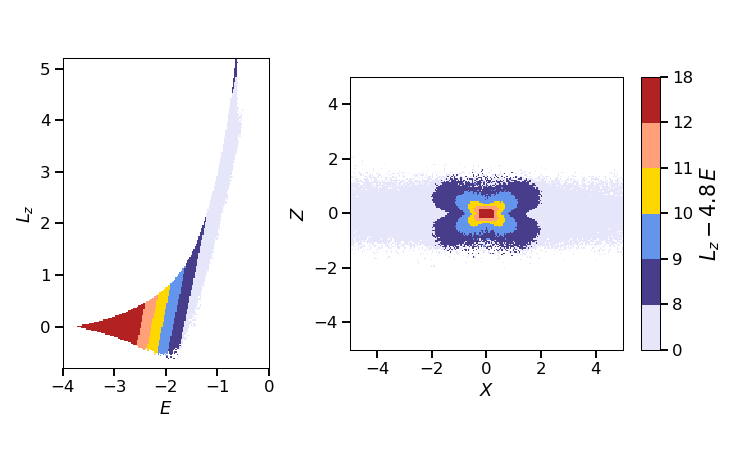}
\end{minipage}
\hspace{0.5cm}
\begin{minipage}[t]{0.48 \textwidth}
\includegraphics[width=\textwidth]{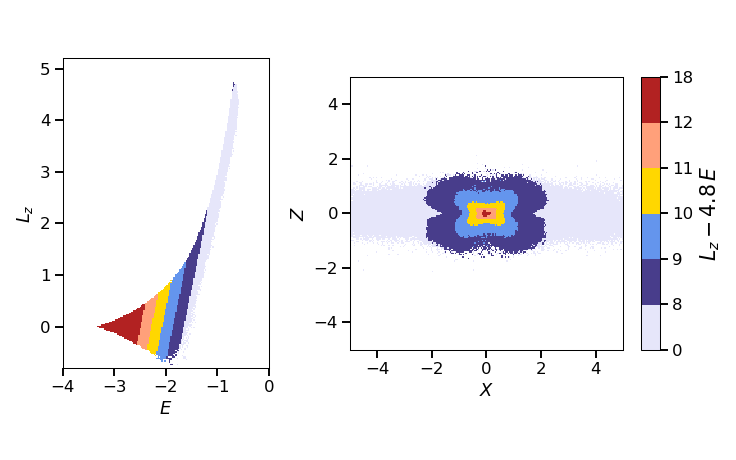}
\end{minipage}
\caption{The distribution of particles on the plane
$E - L_z$ (\textit{left}) and on the plane XZ (\textit{right}). The color indicates the value of the parameter $L_z - 4.8E$; the model SX --- \textit{two plots at left}, the model DX --- \textit{two plots at right}.}
\label{fig:ELz_zones}
\end{figure*}

According to the parameter $E_j$ ($C_j$), we divide the B/PS bulge into six regions (zones). The division is presented in the Table~\ref{tab:ELz_zones}. Fig.~\ref{fig:ELz_zones} shows how our regions are distributed across the $E-L_z$ plane (left plots) and the XZ plane (right plots) for both models. As can be seen from the figure, zone V is the most central and gravitationally bound, while the region of the disk is the most extended and distant, and it does not include particles belonging to the 3D bar (B/PS bulge). On average the total energy  of particles from the disk to the region V becomes smaller and, consequently, our division by Jacobi energies leads to the division in coordinates space too (see Fig.~\ref{fig:ELz_zones}, XZ panels). 
\begin{table}
\centering
\begin{tabular}{rrc|rrc}
\hline
zone & $C_j$ & $E_j$ & zone & $C_j$ & $E_j$ \\
\hline
\hline
disk & $< 8$  &  $ > -1.7$ & III & $10 \div 11$& $-2.3 \div (-2.1)$\\
I    & $8 \div 9$ & $-1.9 \div (-1.7)$ & IV  & $11 \div 12$& $-2.5 \div (-2.3)$ \\
II   & $9 \div 10$ & $-2.1 \div (-1.9)$ & V   & $> 12$ & $< -2.5$\\
\hline
\end{tabular}
\caption{Boundaries of zones in SX and DX models according to the constant $C_j$, \textcolor{black}{or the Jacobi energy $E_j$.}}
\label{tab:ELz_zones}
\end{table}
\par
The main benefit of such a division is as follows. We divide all particles into zones according to the Jacobi energy, which is almost conserved during the time interval under consideration. This means that particles collected inside selected zones retain their spatial configuration so we can analyze the orbital composition of each zone and identify the dominance of certain orbital groups from the center to the periphery. 

\subsection{Peanut structure assembly from the center to the periphery}
\label{sec:zones}
\begin{figure*}
\centering
\begin{minipage}{.48\textwidth}
\centering
\includegraphics[width=\textwidth]{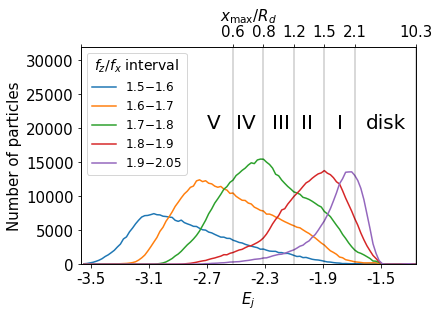}
\end{minipage}
\hfill
\begin{minipage}{.48\textwidth}
\centering
\includegraphics[width=\textwidth]{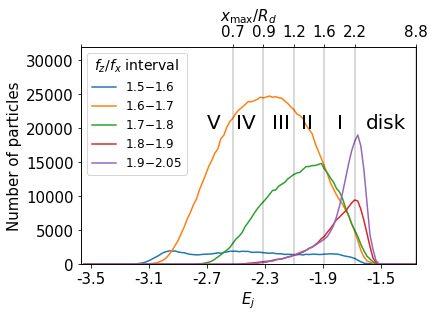}
\end{minipage}
\caption{Distribution of different orbital \textcolor{black}{groups} (see the legend) over \textcolor{black}{the Jacobi energy $E_j$}; \textit{left} --- SX model; \textit{right} --- DX model. \textcolor{black}{The upper scale is the maximal extension of the orbits with the Jacobi energy $<E_j$ along the $x$-axis, expressed in radial scales of the disk, $R_\mathrm{d}$ from Eq.~\eqref{eq:sigma_disk}}.}
\label{fig:Cj}
\end{figure*}

\begin{figure}
\centering
\begin{minipage}{.49\textwidth}
\centering
\includegraphics[width=0.8\textwidth]{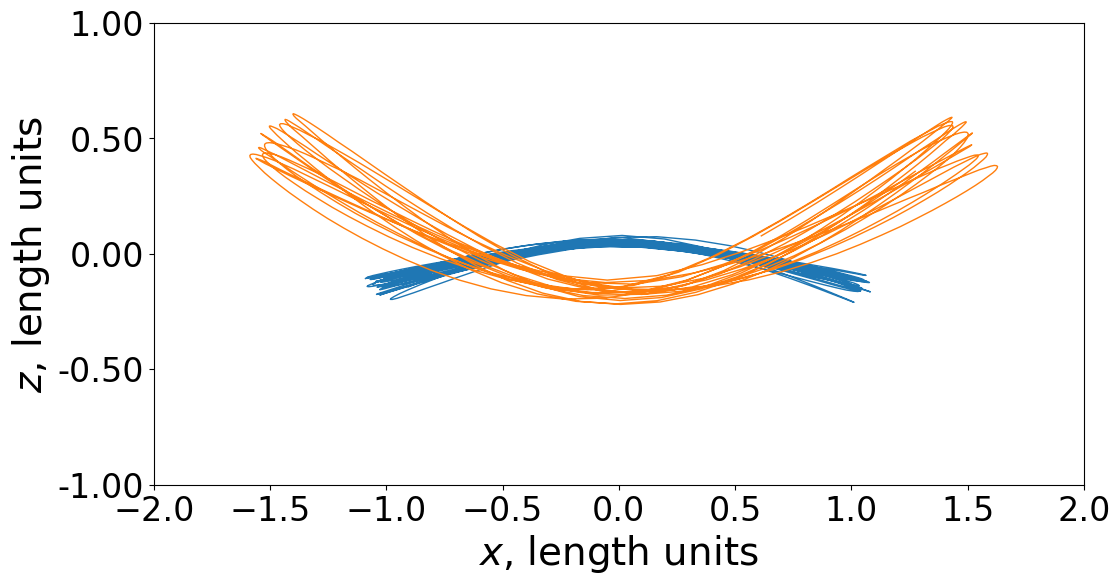}%
\end{minipage}
\caption{Two banana orbits in SX model; \textit{blue line} \textcolor{black}{$E_j=-2.2$}; \textit{yellow line} \textcolor{black}{$E_j=-1.9$}. \textcolor{black}{The orbits are calculated at the time interval $t=400-500$.}}
\label{fig:BAN2}
\end{figure}

Fig.~\ref{fig:Cj} presents the distribution of different groups of orbits over \textcolor{black}{the Jacoby energy $E_j$}. The distribution shows how likely it is to meet the orbit of a given class, \textcolor{black}{defined} by the frequency ratio $f_z/f_x$, in one or another area of the B/PS bulge. Vertical  lines indicate the boundaries of zones inside the B/PS bulge divided according to the \textcolor{black}{Jacoby energy $E_j$ (or constant $C_j$)} (see Fig.~\ref{fig:ELz_zones}). As can be seen from Fig.~\ref{fig:Cj}, each group of orbits has a wide distribution. In the model SX, the orbits with $f_z/f_x<1.6$ tend to inhabit the inner areas, and it is almost impossible to meet such an orbit in a region adjacent to an external disk. On the contrary, `bananas' have a greater \textcolor{black}{extension} on average and are more common in the outer areas of the B/PS bulge. But this is not a strict rule. There are short `bananas' in the inner areas of the bulge (Fig.~\ref{fig:BAN2}). The remaining orbits can be found almost everywhere.
\par
The DX model gives another picture of the distribution over \textcolor{black}{$E_j$} for different groups of orbits. Here the occurrence of orbits in the B/PS bulge has a stricter regularity. The `brezels' are everywhere, but their fraction falls to the periphery, and orbits with the frequency ratio $f_z/f_x>1.8$ are encountered with rare exceptions only at the periphery.
\par
\textcolor{black}{Fig.~\ref{fig:Cj} shows another important and useful fact. The boundary of 2:1 orbits according to the Jacobi energy gives the boundary within which these orbits exist, and hence the position of the vILR. Determining its position in the changing strongly non-axisymmetric potential of the galaxy with a slowly drifting pattern speed is an ambiguous task. At the same time, the $E_j$ boundary for 2:1 orbits naturally separates the bar from the external disk. In addition, it can be seen from these distributions that 2:1 orbits exist in a rather narrow range of energies $E_j$, unlike other orbits.}
\par
The Tables~\ref{tab:Q12_zones}--\ref{tab:Q16_zones} summarize the relative contributions of different types of near-periodic orbits in five zones from Tab.~\ref{tab:ELz_zones} distinguished by the constant \textcolor{black}{$E_j$}.
\par
The general trend in the distribution of the orbits is that orbits with higher ratios of $f_z/f_x$ inhabit the more external areas of the B/PS bulge (I and II zones). At the same time, there are no orbits with a ratio $f_z/f_x=2.0$ in the very central regions zone V). We note, that in each model there are spatial zones where the contributions of different orbital groups are nearly the same. For example, orbits with $f_z/f_x =2.0\pm 0.05$ contribute 40\% in zone I while orbits with $f_z/f_x=1.85\pm 0.05$ give 38\% in the same zone in the model SX. In DX model, a similar situation arises in zone I for orbits with $f_z/f_x=1.65 \pm  0.05$ and $f_z/f_x=1.75 \pm  0.05$.
\par 
In general, the composition of the B/PS bulge from different orbits described above agrees with the findings of \citet{Portail_etal2015b} for their models, but the precise picture seems more complicated than ``all orbital classes are at some radius the main component of the 3D part of the bar'' as in \citet{Portail_etal2015b}. First, there exist spatial zones where the contributions of different orbital classes are very similar.  Secondly, even if there is the main family in a particular zone, other less abundant families contribute to this zone with their own weight which is different for each specific family (see Tab.~\ref{tab:Q12_zones}\ref{tab:Q16_zones}). \textcolor{black}{Thirdly}, the overall picture is highly dependent on the model under consideration. There is a smooth transition of the dominance of each particular orbital class from one zone to another in SX model (see Fig.~\ref{fig:Cj}). In contrast, the DX model demonstrates the dominance of one particular orbital class in all but one of the zones. 
\par

\begin{table}
\centering
\begin{tabular}{c|rrrrrr}
\hline
$f_z/f_x$ &  I     & II     & III    & IV     & V      &  all  \\
\hline
\hline
1.5--1.6 &  1\% &  3\% &  6\% &  9\% & 35\% &  15\% \\
1.6--1.7 &  4\% & 14\% & 22\% & 30\% & 42\% &  26\% \\
1.7--1.8 & 17\% & 30\% & 43\% & 49\% & 21\% &  30\% \\
1.8--1.9 & 38\% & 40\% & 25\% &  9\% &  2\% &  19\% \\
1.9--2.05 & 40\% & 13\% &  5\% &  2\% &  0\% &  10\% \\
\hline
\end{tabular}
\caption{Relative contributions of each orbital group to the formation of each zone of the B/PS bulge. SX model with $Q=1.2$.}
\label{tab:Q12_zones}
\end{table}

\begin{table}
\begin{center}
\begin{tabular}{c|rrrrrr}
\hline
$f_z/f_x$ &  I     & II     & III    & IV     & V   & all \\
\hline
\hline
1.5--1.6 &  3\% &  4\% &  4\% &  6\% & 14\% &   6\% \\
1.6--1.7 & 24\% & 48\% & 62\% & 74\% & 81\% &  56\% \\
1.7--1.8 & 26\% & 36\% & 30\% & 20\% &  4\% &  24\% \\
1.8--1.9 & 19\% &  6\% &  2\% &  1\% &  0\% &   6\% \\
1.9--2.05 & 28\% &  6\% &  2\% &  0\% &  0\% &   8\% \\
\hline
\end{tabular}
\caption{Relative contributions of each orbital group to the formation of each zone of the B/PS bulge. DX model with $Q=1.6$}
\label{tab:Q16_zones}
\end{center}
\end{table}

\section{Individual orbits and the X-structure}

\begin{figure}
\centering
\begin{minipage}{.49\textwidth}
\centering
\includegraphics[width=0.49\textwidth]{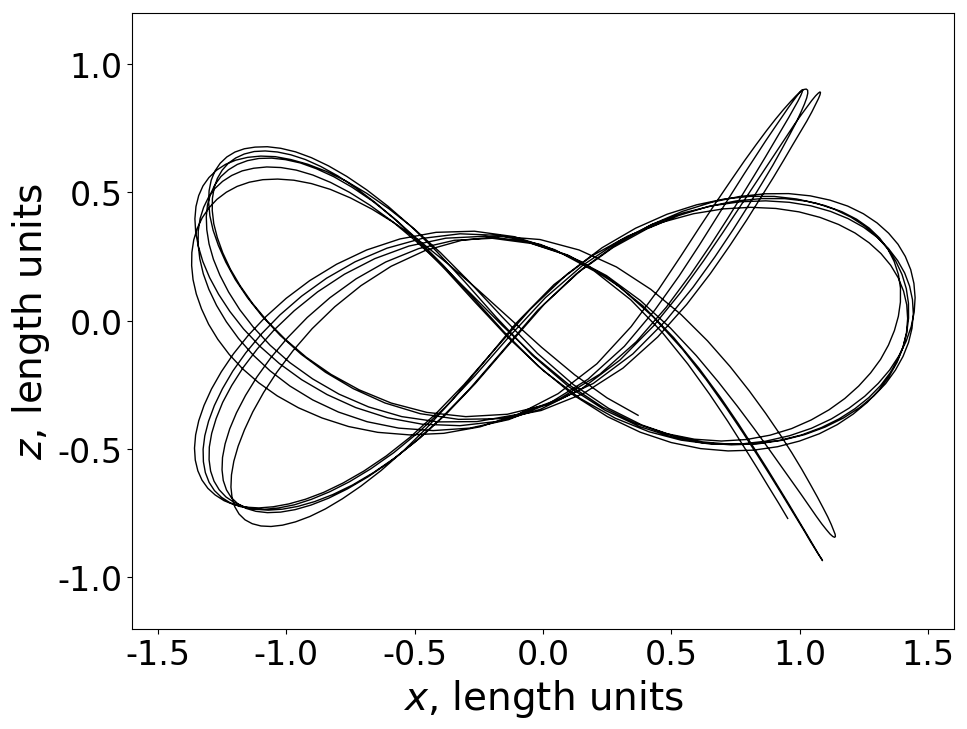}%
\centering
\includegraphics[width=0.49\textwidth]{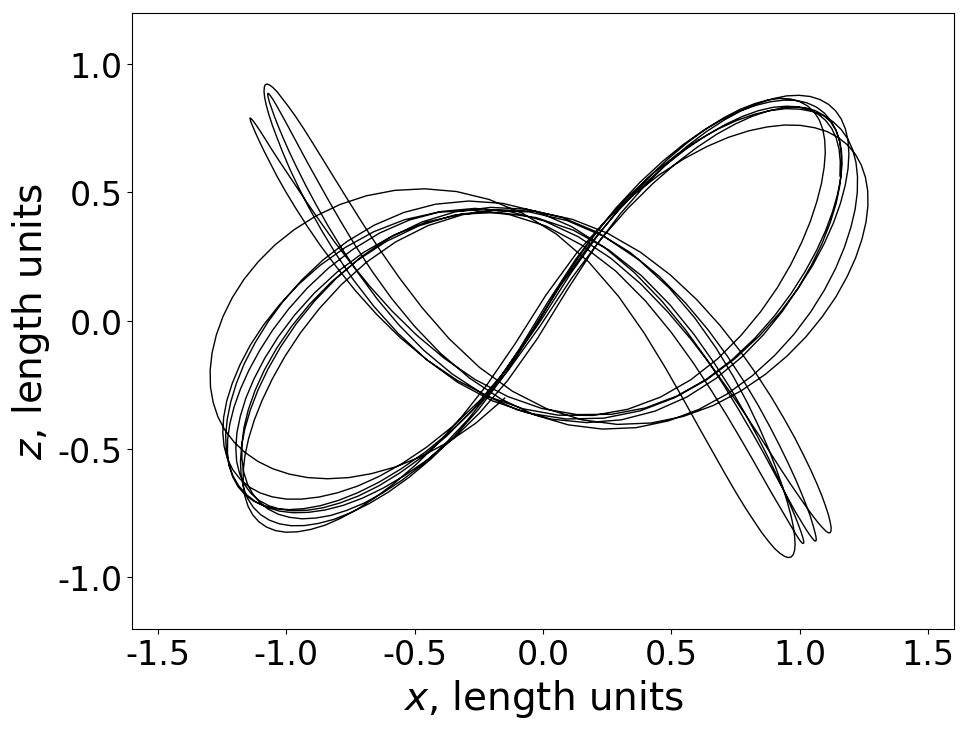}\\
\end{minipage}
\begin{minipage}{.49\textwidth}
\centering
\includegraphics[width=0.49\textwidth]{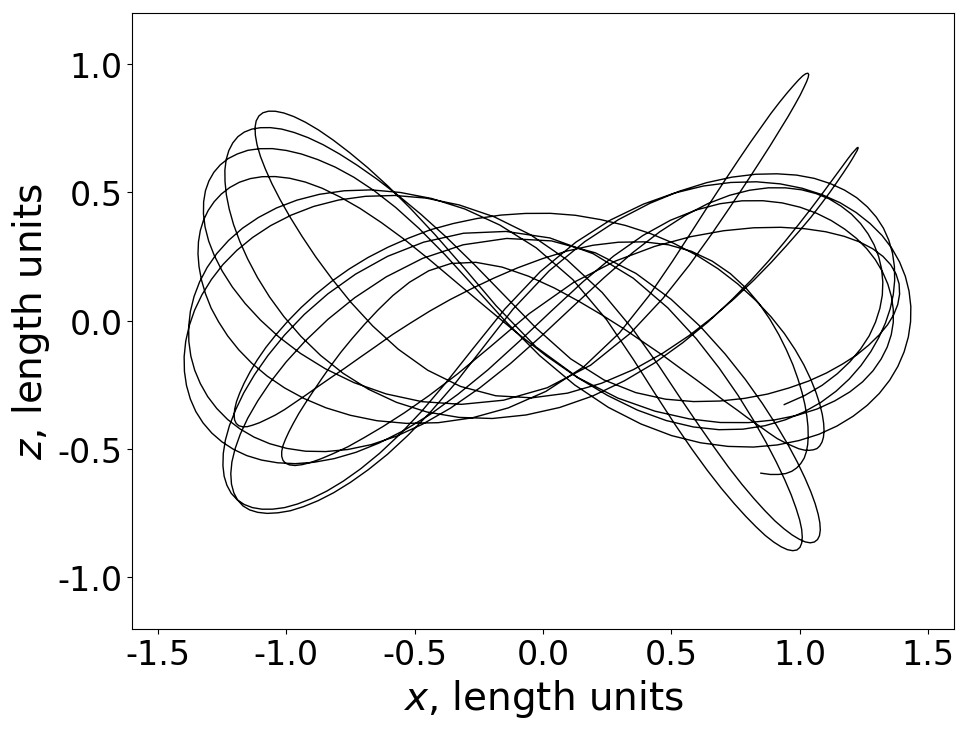}%
\centering
\includegraphics[width=0.49\textwidth]{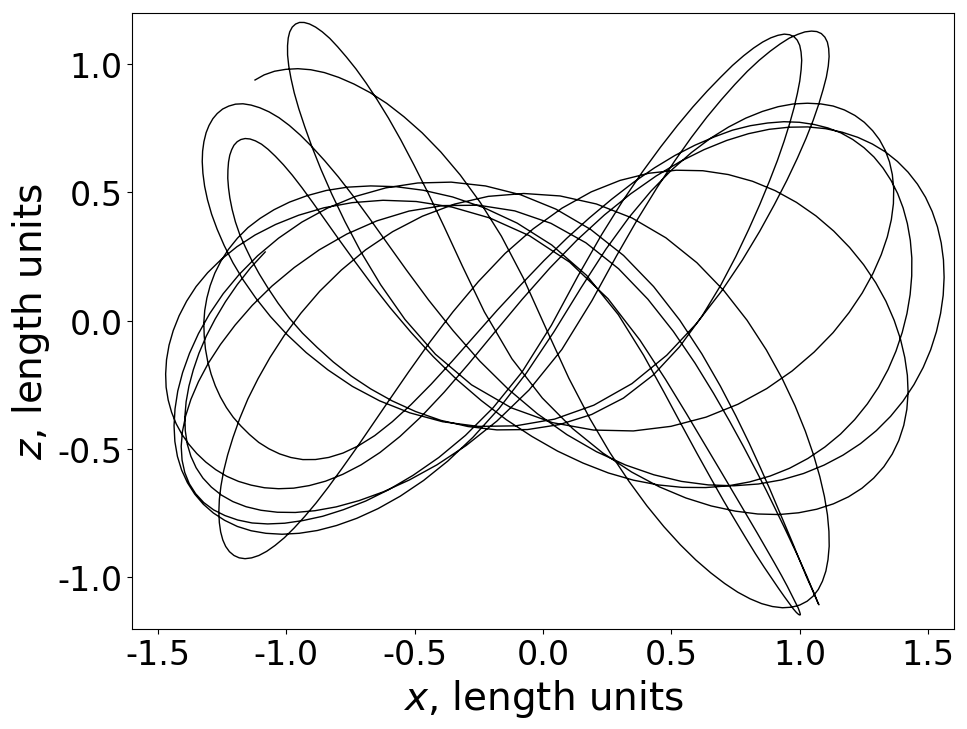}\\
\end{minipage}
\begin{minipage}{.49\textwidth}
\centering
\includegraphics[width=0.49\textwidth]{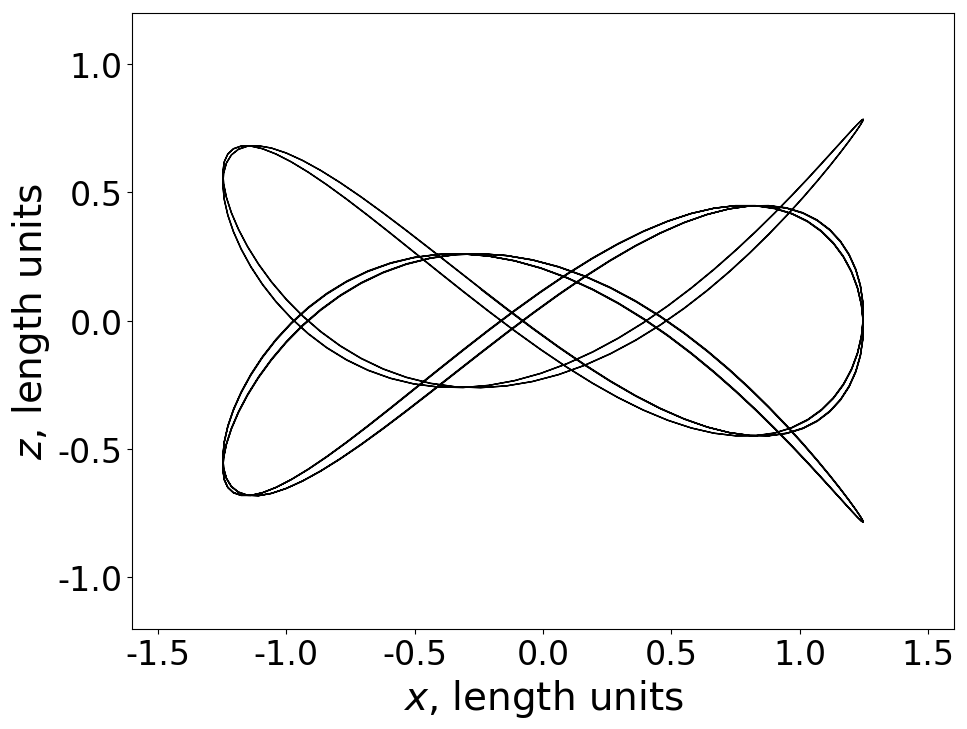}%
\centering
\includegraphics[width=0.49\textwidth]{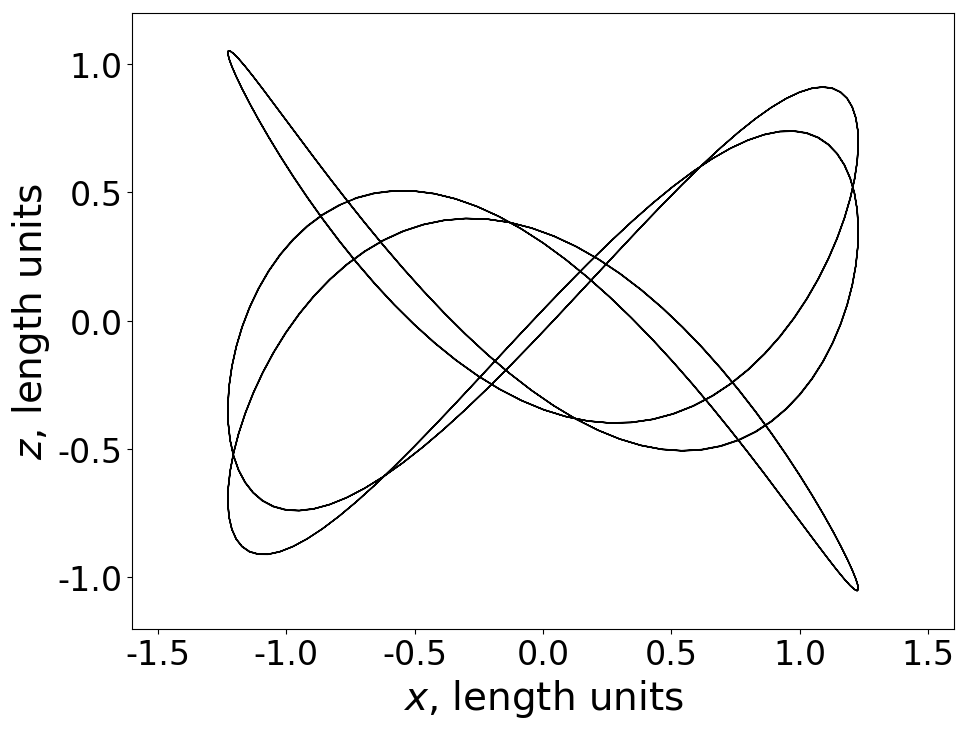}\\
\end{minipage}
\caption{\textcolor{black}{Orbits with $f_z/f_x=7/4$ (\textit{left}) and $f_z/f_x=5/3$ (\textit{right}); \textit{top} --- quasi-periodic orbits; \textit{middle} --- distorted orbits; \textit{bottom} --- parent (`cleaned') orbits of distorted ones. The orbits are calculated at the time interval $t=400-500$.}}
\label{fig:shrimp_brezel}
\end{figure}

\begin{figure}
\begin{minipage}{.49\textwidth}
\centering
\includegraphics[width=0.49\textwidth]{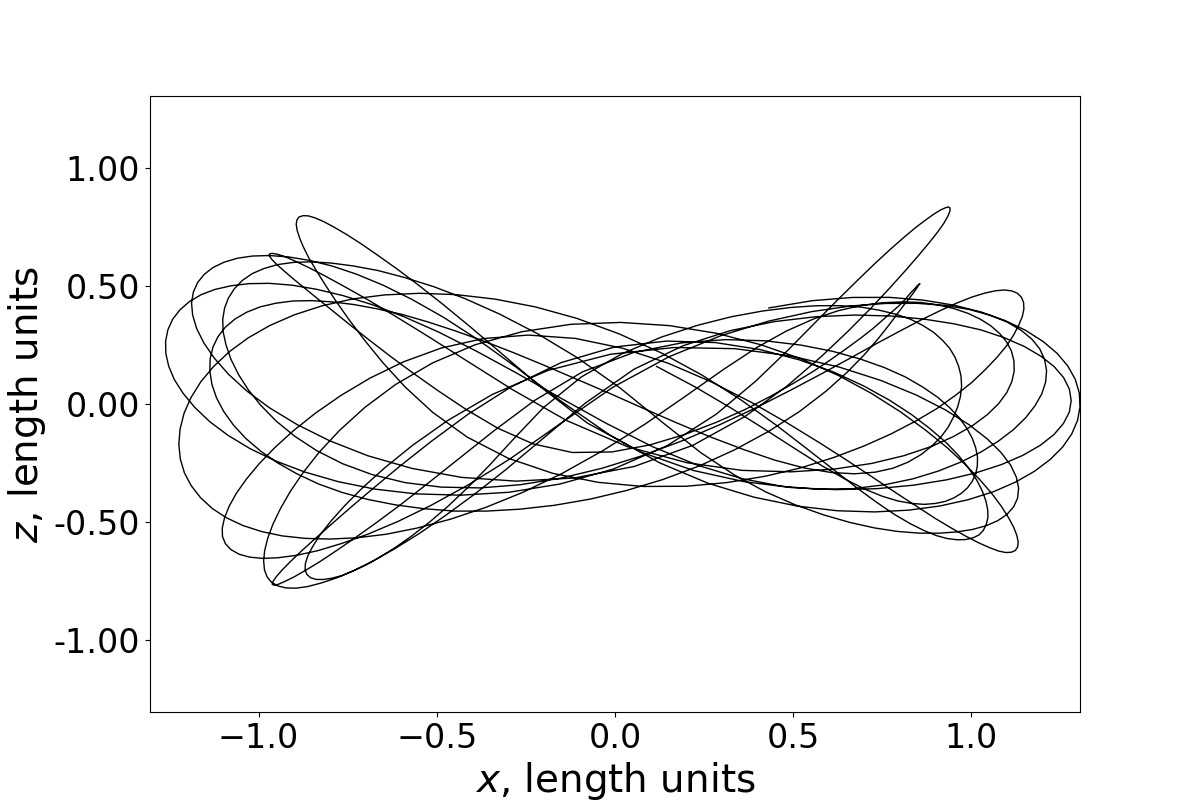}
\includegraphics[width=0.49\textwidth]{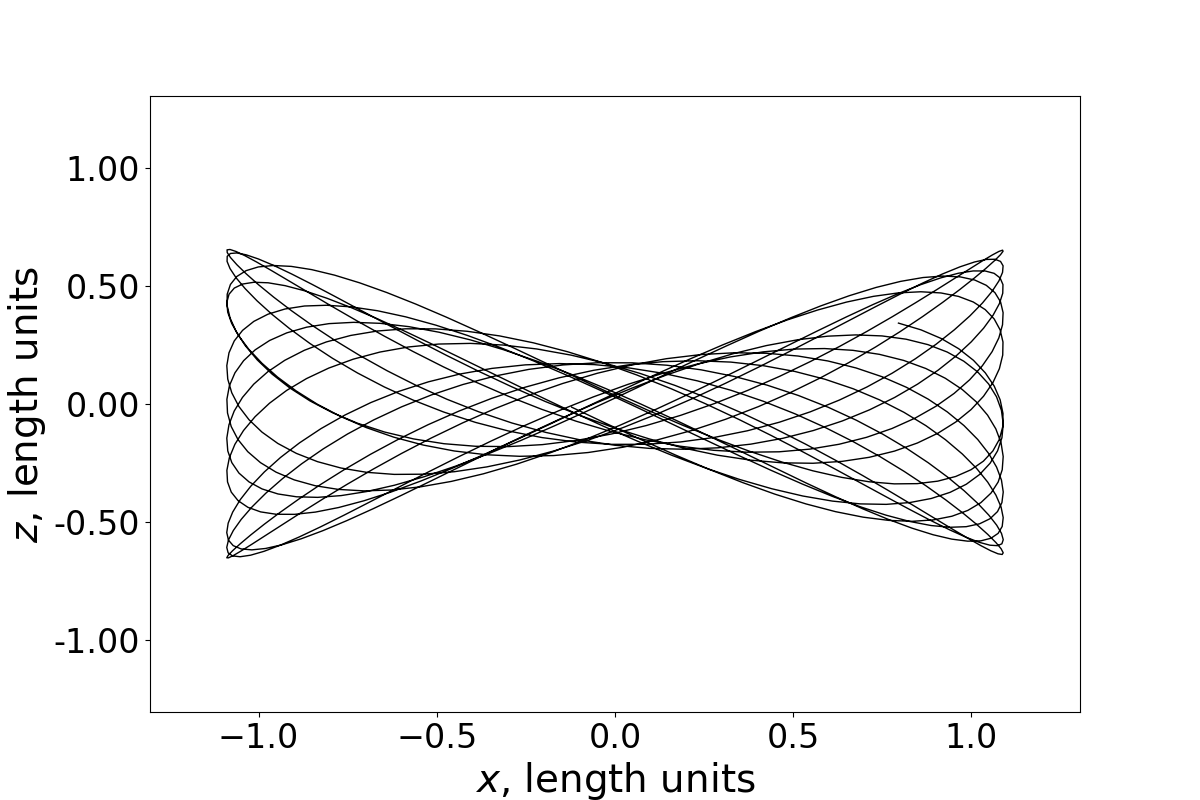}
\end{minipage} 
\caption{\textcolor{black}{Orbits} with $f_z/f_x=1.73$; \textit{left \textcolor{black}{panel}} --- an actual \textcolor{black}{orbit}; \textit{right \textcolor{black}{panel}} --- a parent (`cleaned') orbit. \textcolor{black}{The orbit is calculated at the time interval $t=400-500$.}}
\label{fig:farfalle}
\end{figure}

\begin{figure*}
\begin{minipage}{0.99\textwidth}
\centering
\includegraphics[width=0.33\textwidth]{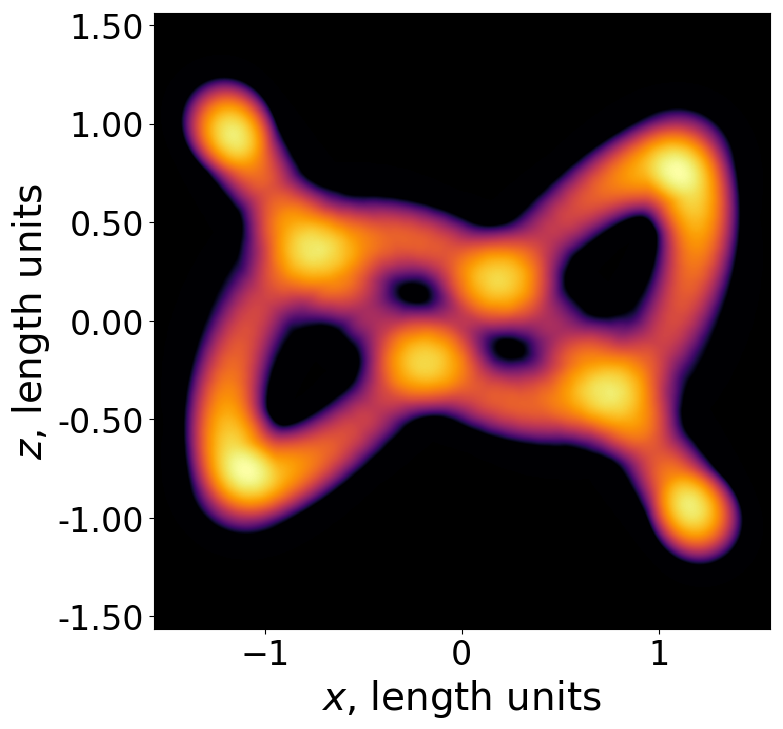}%
\includegraphics[width=0.33\textwidth]{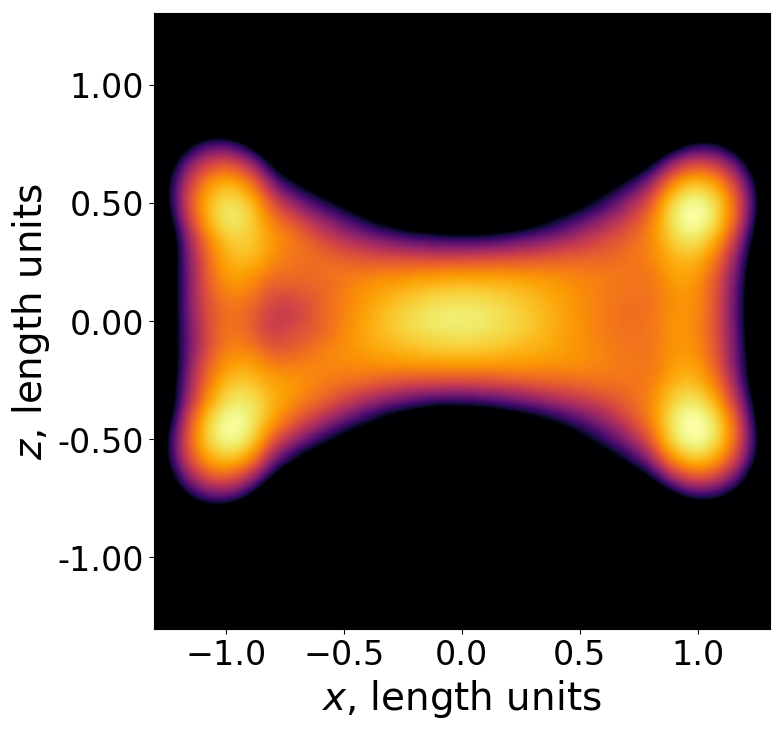}%
\includegraphics[width=0.33\textwidth]{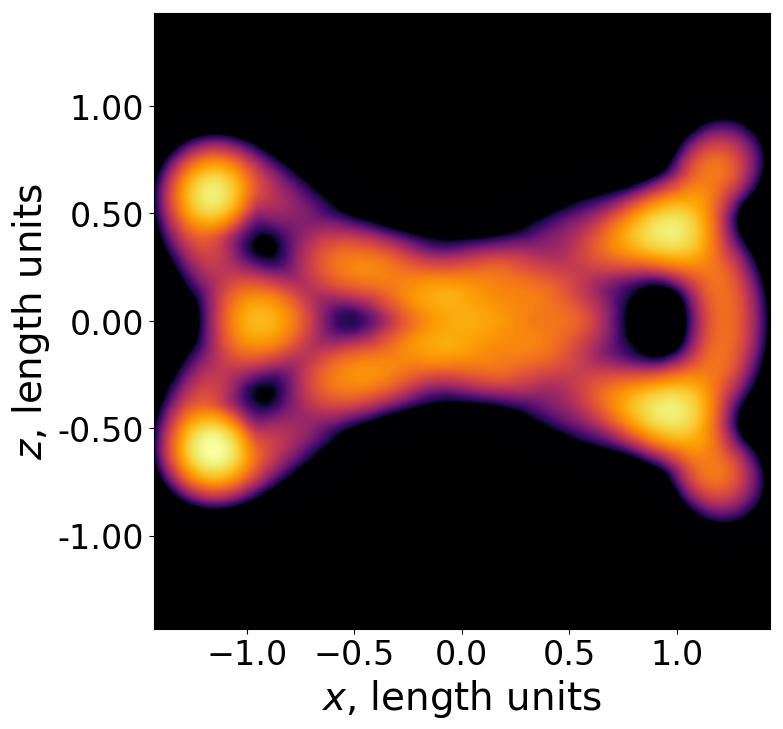}%
\end{minipage} 
\caption{The density distribution produced by an ensemble of `ideal' `brezel' (left), `farfalle' (middle) and `shrimp'-like (right) orbits.}
\label{fig:orbit_dens}
\end{figure*}
Though the orbital composition of the B/PS bulges is already clear, the detailed reasons for the X-structures to appear in our models remain a mystery. \textcolor{black}{\cite{Patsis_Katsanikas2014a} explained the X-structure phenomenon as a consequence of aligning of $z$-maxima in apocenters of the 2:1 orbits in a wide range of Jacoby energy $E_j$. Clearly, such an explanation is not suitable for our models as the contribution of 2:1 orbits in B/PS shape is almost negligible. To understand how the X-structure can, in principle, be assembled from orbits belonging to different orbital classes, which are not banana-shaped orbits,} 
we analyze the most typical orbits in both models and check if there is a possible connection between the individual types of orbits and the general appearance of the X-structure.
\par 

\subsection{The shape of the most typical orbits}
In Fig.~\ref{fig:shrimp_brezel} we plot the most typical orbits with the frequency ratio, which can be expressed by the ratio of small integer numbers. For SX model these are 7:4 orbits, while the \textcolor{black}{B/PS} bulge in the model DX mostly contains 5:3 orbits, or `brezel'-like orbits \citep{Portail_etal2015b}. \textcolor{black}{The top plots depict q.p. orbits of both types. The characteristic shape of periodic orbits can be easily recognized in their thick, q.p. versions. In the middle plots one can see other examples of 7:4 (left) and 5:3 (right) orbits at the same energy $E_j$ as in the top plots. Although} 
the ratio of frequencies for orbits is expressed by the ratio of small integers, the orbits themselves are not closed.
\textcolor{black}{Such orbits can} ``unwind'' because the integration of the equations of motion is not performed in an analytic or ``frozen'' potential, as was done, for example, in \citet{Athanassoula2005,MartinezValpuesta_etal2006,Voglis_etal2007,Wozniak_Michel-Dansac2009,Portail_etal2015b,Valluri_etal2016,Abbott_etal2017}, but in the changing potential of the $N$-body model where the bar pattern speed also alters.
\textcolor{black}{We also cannot exclude that these are not even q.p. orbits, but sticky chaotic ones.} The characteristic portrait of such orbits can still be recognized after some additional processing. In a pair to each real orbit, we plot an ``ideal'' orbit obtained by the processing of an actual orbit. In short, an ``ideal'' orbit is an actual orbit that has only two lines, with a dominant frequency and the second frequency in amplitude, in its periodogram (see Appendix A for details), i.e. an ``ideal'' orbit is just the actual orbit cleared of noisy lines in its $P_x$ and $P_z$ periodograms. 
\par
\textcolor{black}{The closed orbit 5:3 from the right panels of Fig.~\ref{fig:shrimp_brezel} was previously observed in many $N$-body simulations \citep{Portail_etal2015a,Wang_etal2016,Valluri_etal2016,Abbott_etal2017}. It apparently comes from the x1\textit{mul}3 orbit family (see \citet{Patsis_Athanassoula2019}, \textbf{vm33u} orbit).} 
\textcolor{black}{The closed orbit 7:4 from the left panels of Fig.~\ref{fig:shrimp_brezel} has not attracted attention so far.
This orbit has been revealed by \citealt{Patsis_Athanassoula2019} only recently\footnote{\textcolor{black}{It resembles an orbit in figure~8 (bottom, right plot) in \citet{Lokas2019}.}} (see their \textbf{vm41u} orbit), although it may be the parent orbit for many orbits in our SX model.}
\textcolor{black}{The ``ideal'' prototype of the orbit as well as its q.p. version in Fig.~\ref{fig:shrimp_brezel} (left)} resembles a shrimp with a tail and a head with two antennae. For brevity, we will call it a `shrimp'-like orbit.
\par
As can be seen from the Fig.~\ref{fig:fzx}, the $f_z/f_x$ distribution is smooth. According to that, most of the other orbits
have a ratio of frequencies $f_z/f_x$ that falls between simple resonance ratios such as 1.6$=$8:5, 1.667$=$5:3, 1.75$=$7:4 or 1.8$=$9:5. In other words, for most orbits, the $f_z/f_x$ ratio can be expressed within the accuracy of calculations as
an irreducible fraction with large numbers in the numerator and denominator. Fig.~\ref{fig:farfalle} shows the example of such an orbit and its ``ideal'' prototype. 
\textcolor{black}{Even ideal orbits of this type are  non-periodic}
and they fill large areas on the XZ plane, which leads to the creation of the characteristic pattern in the form of a `farfalle', or a`bow tie' pasta.  \textcolor{black}{Perhaps, this type of orbits can be associated with q.p. orbits around plane x1 orbital family~(\citealt{Patsis_Katsanikas2014a}, see their figure 15;~\citealt{Chaves_etal2017,Patsis_Harsoula2018}). Additional studies in terms of Poincare surface sections are necessary to justify the origin of such orbits.}
\par
Presented orbits do not demonstrate a strict X-shape form.  However, at least in the case of a `farfalle'-like orbit, the \textbf{envelope} of the orbit resembles a cross similar to that of the X-structure. This fact can be exploited to reveal the nature of the X-structure \textcolor{black}{in our models}. The key is to consider not only the general pattern of the individual orbit but the distribution of the actual density produced by an ensemble of stars moving along such orbits. 

\subsection{Time averaged orbits}
Suppose that a probe star moves along a `farfalle'-like orbit. The probability of finding a star in different regions of the orbit is different. 
The probability is proportional to the time that the star spends in this region  which, in turn, is inversely proportional to the absolute value of the velocity. The absolute value of velocity changes along the orbit, and there are some extreme points where the minimal value of velocity is reached. \textcolor{black}{Therefore, each particular orbit produces non-uniform density distribution with specific gradients which depend on a particular orbit morphology. The ensemble of each class of orbits assembles into a non-uniform density distribution which depends on both the morphology of orbits of a given group and the abundance of a group. For x1v1 orbits the extreme points are two $z$-maxima, at which there will be the density enhancements. In the model considered by \citealt{Patsis_Katsanikas2014a} the density enhancements produced by such orbits (and their perturbed versions) are aligned into straight segments which we observe as X-structures (see their figure 10). For orbits of a more complicated morphology (Fig.~\ref{fig:shrimp_brezel}) the average density picture should be more complicated.
Multiple intersections or folding of loops of a particle's trajectory also can contribute to the density enhancements along the orbit. Before jumping to the detail studies of the ensembles of different orbital types  we would like to understand what density distribution is produced by individual orbits from Fig.~\ref{fig:shrimp_brezel}~and~\ref{fig:farfalle}}. 
\par 
\textcolor{black}{Fig.~\ref{fig:orbit_dens} depicts the density distribution of typical orbits.}
We took ``ideal'' orbits from Fig.~\ref{fig:shrimp_brezel} and Fig.~\ref{fig:farfalle} and plotted them in such a manner that each point of the orbit is marked by a color which corresponds to the density value in this point. The color \textcolor{black}{scheme} is the same as in Fig.~\ref{fig:xz_res}. The density itself is equal to the number of trajectory points that fall into some fixed neighborhood of the point where the density is calculated. The time step is fixed and is equal to $\Delta t = 0.125$, as in the initial time series. The density is calculated in a circle with a radius of $0.1$ units of length. As can be seen from the figure, the density pattern of a `farfalle'-like orbit indeed has an X-shaped form \textcolor{black}{but the brightest features of the averaged orbit are four prominent clumps at $z$-maxima of the orbit.}
The cases of `brezel' and `shrimp'-like orbits turn out to be more complicated. The density distribution of `brezels' does not show X-shape morphology, but there are some bright spots that cannot produce an X-shape form on their own. The brightest (and densest) clumps on such an orbit roughly correspond to the extreme points of the loops of the orbit. As for the `shrimps', they can give an internal X-shaped structure with their mirror-symmetric counterparts, \textcolor{black}{and they also show bright spots on the periphery \textcolor{black}{at $z$-maxima of the orbit}}. 

\subsection{Understanding the X-structure}

\begin{figure*}
\begin{minipage}{0.99\textwidth}
\centering
\includegraphics[width=0.49\textwidth]{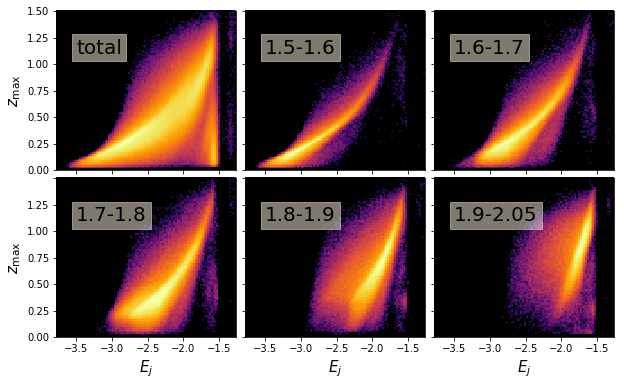}%
\includegraphics[width=0.49\textwidth]{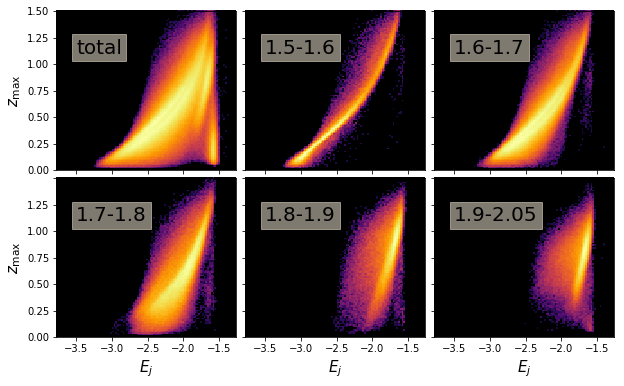}%
\end{minipage} 
 \caption{2D distribution of orbits over the \textcolor{black}{energy $E_j$} and the maximal distance to the disk plane $z_{\mathrm{max}}$ for SX (\textit{left}) and DX (\textit{right}) models. Individual subplots after the first one at top left correspond to different groups of orbits with particular values of the ratio $f_z/f_x$. The values are indicated in legends.}
\label{fig:z_cj}
\end{figure*}
The X-structure seen in Fig.~\ref{fig:XY_XZ} looks mostly like a smooth and unified structure across large spatial areas. Even if we assume that the X-structure is associated with a certain group of orbits (like `farfalle' orbits), the question is how can the orbits in different spatial zones (but still within the same group of orbits selected by the frequency ratio) produce segments of the X-structure, which are smoothly adjacent to each other. To identify the \textcolor{black}{reasons for this behaviour}, we consider the distribution of stars in terms of integral parameters of their orbits. We use the parameters which characterize the extreme points of each individual orbit, the Jacobi constant and the maximal distance of the orbit from the disk plane. The benefits of the usage of these parameters will become clear below. 
\par
Fig.~\ref{fig:z_cj} shows the 2D distribution of orbits in these coordinates for both models for different groups of orbits with a particular value of the ratio $f_z/f_x$. 
\textcolor{black}{In contrast to Fig.~\ref{fig:orbit_dens}, which demonstrates the time averaged ideal orbits, and even periodic ones, $z$-maxima in Fig.~\ref{fig:z_cj} are calculated for all orbits from a given group (possible periodic orbits, quasi-periodic ones, and even, possibly, chaotic orbits).}
While the distribution of each group covers large areas of the parameter space, there is a clearly distinguishable strip, which is the site of an increased density of orbits. \textcolor{black}{We note that part of the orbits has $z$-maxima near zero (flat orbits) and there are orbits that lie above a bright strip. But, as can be seen from the figure,} such a strip of density enhancement exists for each particular orbital \textcolor{black}{family, including orbits with $f_z/f_x \approx 2.0$.} It means that extreme points of most of the orbits within one orbital family tend to line up along some smooth curve\footnote{We did not study this question deeply, but we can assume that a bright strip is formed by the $z$-maxima of q.p. orbits and their distorted counterparts, otherwise it is difficult to explain such a regular behavior.}. As shown in Fig.~\ref{fig:orbit_dens}, the extreme points of the orbits also correspond to the brightest spots of the density distribution produced by an ensemble of each particular class of orbits. We note, that the last statement is true not only for `farfalle'-like orbits but also for `brezels' and `shrimp'-like orbits. 
\par 
\textcolor{black}{Thereby} we obtain that \textit{a)} stars tend to spend
more time in extreme points of loops of the orbit and \textit{b)} these extreme points line up along some smooth curve in spatial coordinates for each particular group of orbits with the fixed ratio $f_z/f_x$. 
The gradient of these curves and their extent along the abscissa are different for different groups of orbits. Each curve composed of density clumps at the extreme points of the orbit within a separate orbital group gives its own ray of the X-structure. But together they merge into a certain averaged ray, \textcolor{black}{with the geometric parameters determined by the dominant orbital group in the general ensemble of orbits.}
\par 
\textcolor{black}{Figure~11 in \citet{Patsis_Katsanikas2014a} demonstrates the same phenomenon, but only for periodic x1v1 orbits and their perturbed counterparts, including quasi-periodic orbits trapped by stable x1v1 periodic orbits or even chaotic orbits. The slope of the bright strip and the alignment of $z$-maxima along the line only for large $E_j$ indicate that these orbits contribute to the X-structure only at the periphery. In this case, the ray formed by these orbits begins not from the center (that is, small values of $E_j$), but far from it. This is exactly what we see in Fig.~\ref{fig:z_cj} for 2:1 orbits.}
\par
\textcolor{black}{
Our analysis leads us to the following interpretation of the X-structure phenomenon. An X-structure seems to be a dynamical effect due to non-uniform density distribution produced by each particular orbit and a collective effect due to the fact that these density distributions are summed up in a certain way.  The key distinction from the previous paradigm exploited by \citet{Patsis_Katsanikas2014a} is that each type of orbits actually observed in our models can be used to form the X-shape feature. But the concept that X-structures are \textcolor{black}{density enhancements produced by the aligned up $z$-maxima of orbits interestingly remains the same.}} 
\section{How the X-structure manifests itself for different groups of orbits} 

\subsection{Extracting the X-structure}
In this section, we \textcolor{black}{analyze} properties of X-structures assembled from different orbital groups more accurately by constructing unsharp-masked images for each model. Unsharp-masking is a common technique for extracting X-structures in real galaxies \citep{Bureau_etal2006}. It allows enhancing the local image contrast where there is an abrupt change in color gradations (particle density). In this method, the masked image is obtained by replacing each pixel by the difference between it and the median value in the circular aperture. The radius of the aperture was chosen so as to emphasize the interesting features of the image. A Gaussian filter was also applied to the masked image to smooth out artifacts arising from the subtraction of images. 

\subsection{Orbital ingredients for the X-structure}

\begin{figure*}
\centering
\begin{minipage}{.4\textwidth}
\centering
\includegraphics[width=\textwidth]{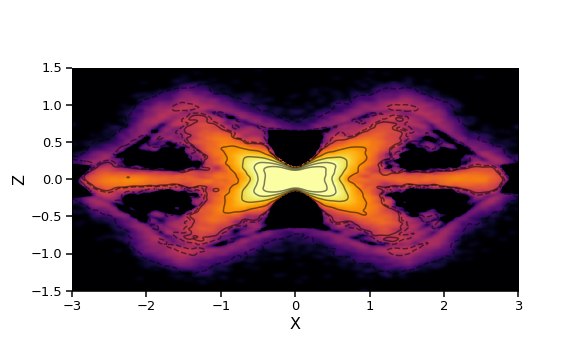}\\
\centering
\includegraphics[width=\textwidth]{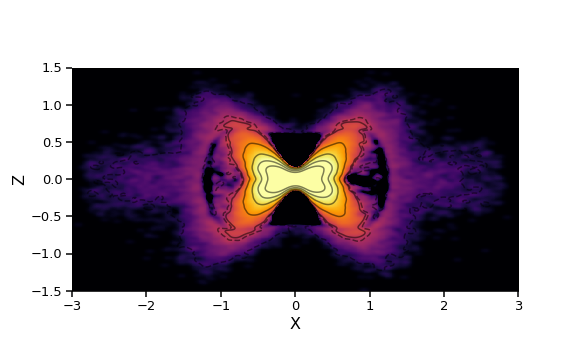}\\
\centering
\includegraphics[width=\textwidth]{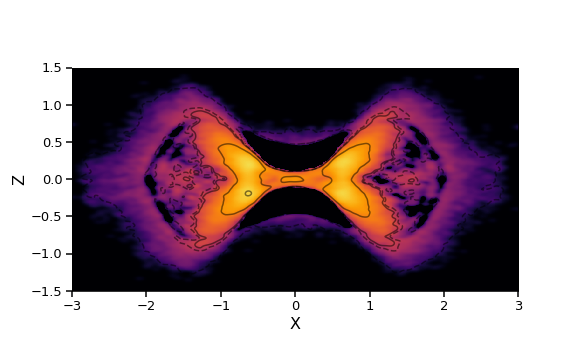}
\centering
\includegraphics[width=\textwidth]{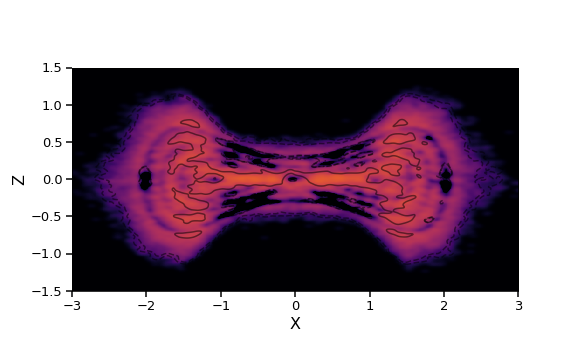}
\end{minipage}
\hfill
\begin{minipage}{.4\textwidth}
\centering
\includegraphics[width=\textwidth]{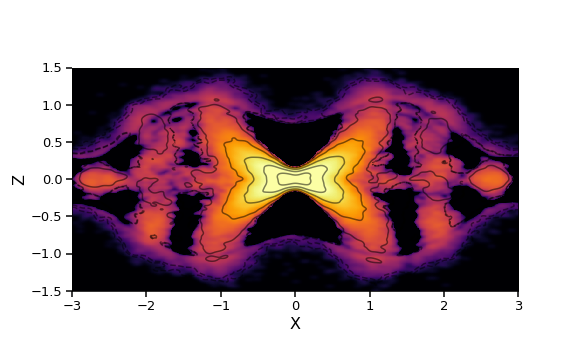}\\
\centering
\includegraphics[width=\textwidth]{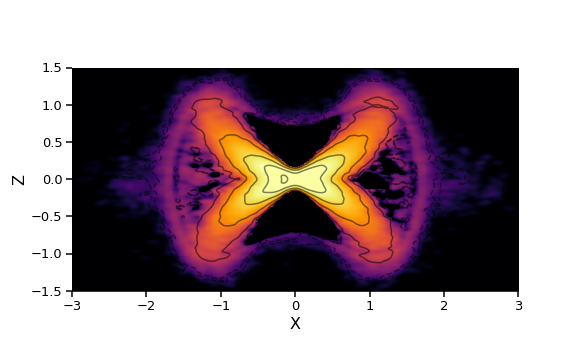}\\
\centering
\includegraphics[width=\textwidth]{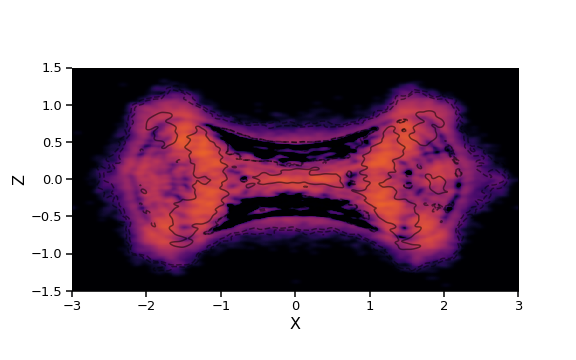}
\centering
\includegraphics[width=\textwidth]{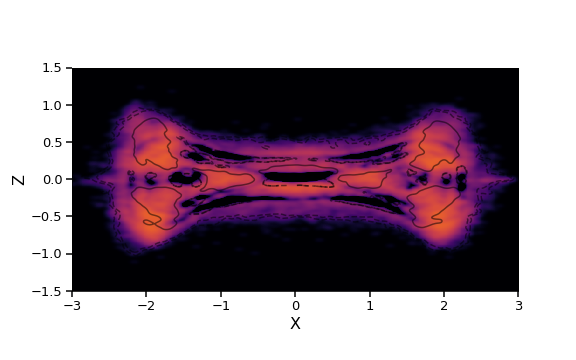}
\end{minipage}
\caption{Unsharp-masking images for SX (\textit{left}) and DX (\textit{right}) models. Projection on the XZ plane. From top to down: all B/PS particles, $f_z/f_x = 1.55-1.75$, $f_z/f_x = 1.75-1.95$ and $f_z/f_x = 1.95-2.05$.}
\label{fig:mask}
\end{figure*}

Masked images for both models for all snapshots from Fig.~\ref{fig:xz_res} are shown in Fig.~\ref{fig:mask}. One can see that all orbits from different parts of the distribution over the frequency ratio $f_z/f_x$ create their own X-structures but with quite different morphology, with the exception of the orbital group with $f_z/f_x = 2.0$ in SX model. The combined X-structure (Fig.~\ref{fig:mask}, top row) is a result of non-linear contributions \textcolor{black}{to the masked image} of each groups of orbits.
\par 
A change in morphology from low to high frequency ratios has a similar pattern as was discussed in the section regarding the B/PS bulge structure. For both models, the X-structure rays that are created from orbits with low frequency ratios $1.55-1.75$ (Fig.~\ref{fig:mask}, second row from the top) look shorter and thicker than rays in other plots. They have a large opening angle with almost no bridge between them. With an increase in the frequency ratio ($1.75-2.05$), the X-structure becomes more and more flattened, more radially extended and the central bridge between the rays increases its size.
\par 
X-structures that are produced by the same groups of orbits also differ between the two models. These differences could already be foreseen in snapshots in Fig.~\ref{fig:xz_res} but the unsharp-masking procedure outlines them more clearly. First, the X-structure rays of the low frequency ratios ($1.55-1.75$) in SX model are less elongated in both vertical and radial directions than those in DX model (Fig.~\ref{fig:mask}, second row from the top). This is because the corresponding orbits in DX model are not only more numerous (Fig.~\ref{fig:fzx}), but also have a wider habitat. According to Fig.~\ref{fig:Cj}, the corresponding orbits in DX model are distributed in a much wider range of the \textcolor{black}{energies $E_j$} than analogous orbits in SX model. This means that the length of the region inhabited by these orbits will be greater, although the shape of the region itself and the opening angle of the rays will be the same as in SX model. 
\par
X-structures for higher frequencies ratios also appear quite different in both models. But in this case, the difference mainly arises from the fact that the groups of orbits differ in number in the corresponding models. In particular, orbits with $f_z/f_x = 1.75-1.95$ are more abundant in SX model. Their number is comparable to orbits with $f_z/f_x=1.55-1.75$ and they create very bright rays. The corresponding usharped-masked image is brighter and sharper than its counterpart in DX model. As for the orbits with $f_z/f_x=1.95-2.05$, there are two main differences between the two models. First, there is an empty space between the crescents in the most central region in DX model, which is due to the prevalence of \textcolor{black}{almost periodic} banana-shaped orbits over ``librating'' ones (see section~\ref{sec:v_res}). Secondly, the overall structure in this part of the distribution looks more elongated in DX model. In SX model, image processing leaves only two faint half-shells on the periphery without crescents. \textcolor{black}{This is similar to what is found by  \citet{Abbott_etal2017} (see their figure~7, 2nd row left).}
\par 
The combined image, with the contribution from all ensembles of orbits, turns out to be different for the considered models. In SX model, where the orbits with $f_z/f_x=1.55-1.75$ and $f_z/f_x=1.75-1.95 $ are equally numerous, the opening angle of the combined X-structure is a compromise between the angle given by the orbits with $f_z/f_x=1.55-1.75$, and the angle given by the orbits with $f_z/f_x=1.75-1.95$. As a result, it is less than in DX model, where the main component of the X-structure is $f_z/f_x\simeq1.65$ orbits. 
Since both groups of orbits with $f_z/f_x=1.55-1.75$ and $f_z/f_x=1.75-1.95$ in SX model are equally numerous, but their X-structures give different ray lengths and different opening angles, the combined X-structure has a curved ray.
In the combined masked image such a fracture of the ray is clearly visible in the central areas where the transition from one orbital group to another occurs (Fig.~\ref{fig:mask}, top left plots).
\par 
Whether such a transition occurs in real observed galaxies has yet to be verified by a detailed analysis of the curvature of the corresponding X-structure rays. \textcolor{black}{Although, some X-structures extracted, for example, by \citet{Laurikainen_Salo2017} (see their figure~B4, \textcolor{black}{NGC~5353}) or by \citet{Savchenko_etal2017} (see their figure~4) demonstrate a high degree of curvature, it has yet to be understood whether such an effect is a consequence of real transition of the opening angle or \textcolor{black}{it is due to the} employed photometric models and properties of processing algorithm. 
We also note that this effect is not uniquely specific for our SX model but also observed in numerical models considered by \citet{Portail_etal2015b} (although it did not attract much attention there). Indeed, if one looks closely in figure~3 in \citet{Portail_etal2015b}, the violation of the straightness of the X-ray is well manifested in the central areas of the masked image.}
\par 
DX model demonstrates the opposite morphology. 
\textcolor{black}{Transition here is more abrupt and manifests itself in the presence of additional density enhancements distinguishable from the main X-structure.} 
In principle, such secondary density peaks could be an indicator of the frequency ratio distribution with two distinct humps as in Fig.~\ref{fig:fzx} but in reality, projection effects can well hide such an additional X-structure. This can be seen from the work by \citet{Portail_etal2015b}. Authors give the distribution over the ratio $f_z/f_x$ for `Milky Way'-like models that are very similar to the distribution of our DX model. There is a sharp peak near $f_z/f_x = 5/3$ with a wide plateau from 1.8 to 2.0. Unfortunately, the orbital composition of B/PS bulges for these models was not investigated, but we are sure that such models should show double X-structures. A visual inspection of side-on snapshots of these models in \citet{Portail_etal2015a} shows that there is a slight hint of the double X-structure, but the bar in \citet{Portail_etal2015a} is turned so that the second X-structure in the edge-on projection will be hidden.
\par
Another feature of the structures collected from orbits with different ratio $f_z/f_x$ is that they can produce both intersecting rays on masked images, and rays with bridges. Such morphology of X-structures is observed in galaxies \citep{Bureau_etal2006}. The rays are either CX (centered) or OX (off-centered), depending on whether or not they cross the center of the galaxy (CX) or not (OX). \citet{Patsis_Harsoula2018} focused on two types of 2:1 orbits. They showed that x1v1 orbits (in notation by \citealp{Skokos_etal2002}, or BAN orbits in notation by \citealp{Pfenniger_Friedli1991}) give X-structures of the OX type, and x1v2 (ABAN) orbits lead to the formation of the X-structures of CX type. In our scheme OX rays are characteristic of high values of the ratio $f_z/f_x$ and CX rays are obtained from orbits with low values of the frequency ratio. The cumulative X-structure will have either intersecting in the center or non-intersecting rays, depending on the type of the dominant orbits in the `cooked' B/PS bulge. 
\par
\section{Summary and conclusions}
In the present work, \textcolor{black}{we present models that
show how X-structures can be assembled} 
from orbits with different ratio of vertical to in-plane frequencies $f_z/f_x$.
\par
We analyzed the orbital structure of two self-consistent $N$-body models by means of Fourier analysis of spatial coordinates of each star-particle. The initial conditions for the models differ only in the degree of initial dynamic heating of the stellar disk in its plane, i.e. we consider models with different initial values of the Toomre parameter. In the course of evolution, both models demonstrate the formation of the B/PS bulge with an X-structure; however, in a hotter model, the X-structure turns out to be double.
\par
In contrast to most previous works \citet{Athanassoula2005,MartinezValpuesta_etal2006,Voglis_etal2007,Wozniak_Michel-Dansac2009,Portail_etal2015b,Valluri_etal2016,Abbott_etal2017}, we carried out the frequency analysis in the ongoing $N$-body potential of each model. We considered the time interval well after the bar formation time, $t=400-500$ in our time units, or $5.9-7.4$ Gyr. We determined the values of dominant frequencies $f_\mathrm{R}$, $f_x$ and $f_z$ for all particles of the stellar disk ($N_\mathrm{d}=4 \cdot 10^6$) without exception for both models over this time interval. We found that the usual FFT frequency resolution is insufficient for reliable peak frequency identification in our case. We improved the accuracy of the peak identification using DFT of the time series in the vicinity of the maxima obtained using FFT transform. The resulting accuracy was equal to $\Delta f = 10^{-3}$ in our \textcolor{black}{inverse} time units or $\Delta \omega=0.45$~km/c/kpc. This resolution is several times better than that in \citet{Gajda_2016} in the same time interval. We roughly estimated the line width and obtained that most of the peaks were traced with at least ten data points with this frequency resolution.
\par
With such accuracy, we were able to distinguish between different families of orbits according to the ratio of frequencies, both in the plane (in terms of the ratio $f_\mathrm{R}/f_x$) and in the vertical direction (in terms of the ratio $f_z/f_x$). In contrast to the model considered by \citet{Gajda_2016}, where the bar pattern speed was constant over time, in our models the bars gradually slow down. Despite this difference, we obtained a generally similar picture of frequency distributions. 
\par 
\textcolor{black}{For both our models the distribution over the ratio $f_\mathrm{R}/f_x$ shows the large peak at 2:1 frequency ratio. These particles constitute the bar observed in our models. The distributions also show a family associated with the outer disk and several small families which mostly consist of particles that are not elongated along the bar and inhabit the very central regions of the disk. The bar particles with 2:1 frequency ratio $f_\mathrm{R}/f_x$ were further investigated as the main constituents of the B/PS bulge, part of which is the X-structure.}
\par 
For these particles we examined the distribution over the ratio of the vertical to $x$-axis oscillation frequencies, $f_z/f_x$. In both models the distributions continuously span the region from $f_z/f_x \approx 1.5$ to $f_z/f_x \approx 2.0$. The distribution appears to be smooth in both models without sharp resonance peaks, with the exception of 2:1 orbits.
\par 
Using the Jacobi constant we analyzed how different orbital groups, which are represented in our models, fill various spatial regions of the modelled galaxies. \citet{Portail_etal2015b} stated that ``all orbital classes are at some radius the main component of the 3D part of the bar''. For our models, this statement is valid only in the case of SX model. But even there, each particular family of orbits has a wide spatial distribution, and in some regions it is hard \textcolor{black}{to decide on} a leading family. For our hotter model DX, this concept is just not valid. 
Here, one family of orbits prevails over all others in almost the entire area of the bar, except for the most central and peripheral areas.
We can conclude that in different models there is no unique pattern of the spatial distribution of different orbital families.
\par
 In \textcolor{black}{some works
 \citep{Combes_etal1990,Patsis_etal2002b,Pfenniger_Friedli1991,Quillen2002,Quillen_etal2014,Patsis_Katsanikas2014a,Patsis_Harsoula2018}}
 banana orbits with $f_z/f_x \simeq 2.0$ were considered as the main component of the orbital composition of the B/PS bulges and their X-structures. 
 \textcolor{black}{
 For our models we found that banana-shaped orbits are far too small in number in comparison with other orbital families.} Visually, they are almost \textcolor{black}{invisible against} the background of the vast number of orbits with lower ratios of $f_z/f_x$.
  \par 
 Nevertheless, bright X-structures appear in both our models. To resolve the issue, we considered different slices of $f_z/f_x$ distribution from $1.5$ to $2.0$. We obtained that each of them appears as sub-part of the X-structure in a form of its own X-structure seen side-on. This effect is especially prominent in the unsharp-masked images of each of such slices. For a detailed explanation of this, we considered density distributions produced by the most typical orbits in each of our models. In DX model, these are `brezel'-like orbits introduced by \citet{Portail_etal2015b} with $f_z/f_x = 5/3$, while in SX model orbits with $f_z/f_x= 7/4$ prevail. The characteristic pattern of an ideal prototype of such an orbit has two antennae and tail, and we call such orbits `shrimps'. Since the entire distribution appears as a smooth function of $f_z/f_x$, we also considered an intermediate type of orbits with a frequency ratio that cannot be expressed by a simple irreducible fraction. Such orbits look like a `farfalle' and sweep up a B/PS-like area.  We found that the density distribution produced by the ensemble of stars moving along exactly the same orbits of one of the types has a non-uniform distribution with the brightest points corresponding to the extreme points of the loops of such orbits. Considering the spatial distribution of such extreme points for different orbital groups, we found that within one group, extreme points are mostly line up along a smooth curve fixed for each particular group. This led us to the following interpretation of the X-structure phenomenon. We can conclude that the X-structure 
 is a place of concentration of loops (caustics) of various orbits and, accordingly, a place of the increased particle density. \textcolor{black}{\citealt{Patsis_etal2002b,Patsis_Katsanikas2014a} explained the observed X-structures in a similar manner but by means of 2:1 orbits with a small mixture of specific orbits of other types~\citep{Patsis_Katsanikas2014a, Patsis_Harsoula2018, Patsis_Athanassoula2019}. One of the primary results of the present work is that orbits from other groups with \textcolor{black}{morphology quite different from} banana-shaped can produce linear segments of the density enhancements too. These segments are observed as an X-shape that encloses the peanut structure produced by such orbits. The open question which has yet to be answered is why the extreme points of the loops are ordered in space in the form of the straight ray for each particular class of orbits.} The whole X-structure is the product of the superposition of such rays. According to that, orbits with different values of the ratio $f_z/f_x$ are involved in the assembly of the X-structure, and X-structures are not uniquely associated with the orbits 2:1. Moreover, in the case of our models, 2:1 orbits are far from the main building material for the X-structure and cannot be its `backbone'.
 \par 

\par
Despite the similarity of the numerical models under study, the present analysis revealed a significant difference in their orbital composition. It results in a prominent difference in the morphology of their B/PS bulges and X-structures. In particular, it turned out that the opening angle of the X-structure is determined by orbits, which dominate in the assembly of the B/PS bulge. 
Orbits with a low frequency ratio $f_z/f_x$ are assembled into an X-structure with a large opening angle. Orbits with a larger frequency ratio give a more flattened and elongated X-structure.
Probably, the observed scatter \citep{Laurikainen_Salo2016, Ciambur_Graham2016,Savchenko_etal2017} of the opening angles of X-structures in real galaxies is associated with this circumstance. At the same time, we came to the conclusion that when there is no dominant family of orbits with respect to the frequency ratio $f_z/f_x$, then the ray of the emerging X-structure can be curved and has a larger opening angle in the center and a smaller one on the periphery. Processing of real galaxies possessing X-structures can reveal how common is this phenomenon. 
\par
We also obtained that the dominance of a certain group of orbits determines another feature of the X-structures, noticed in the observations: the presence of X-structures with intersecting rays and X-structures with bridges. This effect was also demonstrated by~\cite{Patsis_Harsoula2018} but in the framework of 2:1 stable (BAN) and 2:1 unstable (ABAN) orbits. In our models intersecting rays are characteristic of orbits with lower frequency ratios ($f_z/f_x \approx 1.5-1.85$) and bridges arise due to orbits with higher frequency ratios ($ f_z/f_x \gtrapprox 1.85$). 
\par 
Although our results are generally consistent with previous works there are \textcolor{black}{essential} differences due to \textcolor{black}{the discrepancy} between the models under study. \citet{Portail_etal2015b} considered models with $Q=1.4$, that is, dynamically hotter than our SX model with $Q=1.2$. Their M85 model has the distribution over the ratio $f_z/f_x$ with no wide `hump' as in our SX model. At the same time, the frequency ratio distribution for M85 model does not show two separate groups of orbits as in our DX model. It appears that M85 model analyzed in \citet{Portail_etal2015b} is in between our models in terms of B/PS bulge properties. The existence of such in-between cases shows that the relationship between the parameters of the initial galaxy model and the orbital composition of the B/PS bulges is complex, and it has yet to be studied in a larger number of different models.
\par 
In the present work we did not consider the evolution of the X-structure over time. However, previous numerical works \citep{Smirnov_Sotnikova2018} showed that there is a dependence of X-structures parameters on time. Namely, an X-structure tends to increase the length of the rays and decrease their opening angles. As shown in the present work, the dominance of a certain group of orbits leads to different values of opening angles of the X-structure. Consequently, the evolution of the X-structure can be associated with a gradual change in orbital composition which, \textcolor{black}{in turn, may be related to} the resonant capture of new orbits or the transition of orbits from one class to another. This hypothesis has yet to be verified by examining the orbital composition of the model for longer periods of time.

\section*{Acknowledgements}
The authors express gratitude for the grant of the Russian Foundation for Basic Researches number 19-02-00249. 
We also thank Amy Jones for reading the manuscript and improving the language. \textcolor{black}{We are very grateful to the anonymous reviewer for a deep and careful reading of the manuscript and many valuable comments that contributed to a significant improvement of the scientific quality of the manuscript and a clearer presentation of our results.} The authors acknowledge the usage of {\tt Py-SPHViewer}~\citep{py-sphviewer} and {\tt UNSIO} (\href{https://projets.lam.fr/projects/unsio}{https://projets.lam.fr/projects/unsio}) software packages and express gratitude to its authors.

\bibliography{article3_aj_style}

\appendix
\section{Extracting closed orbits}

\begin{figure}
\centering
\begin{minipage}{.23\textwidth}
\centering
\includegraphics[width=\textwidth]{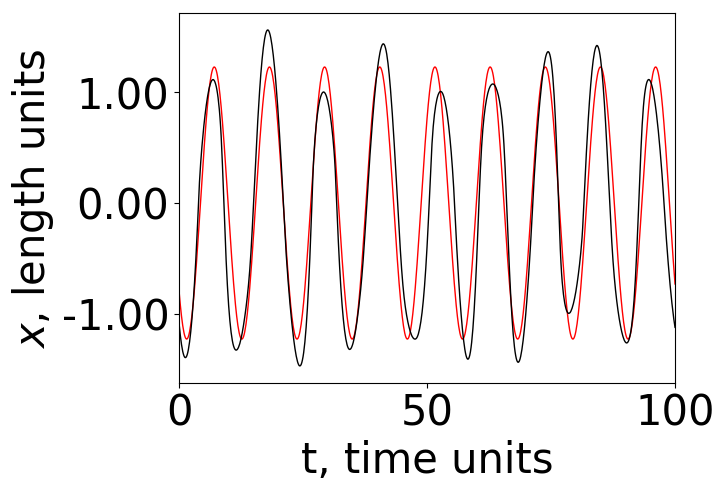}%
\end{minipage}
\begin{minipage}{.23\textwidth}
\centering
\includegraphics[width=\textwidth]{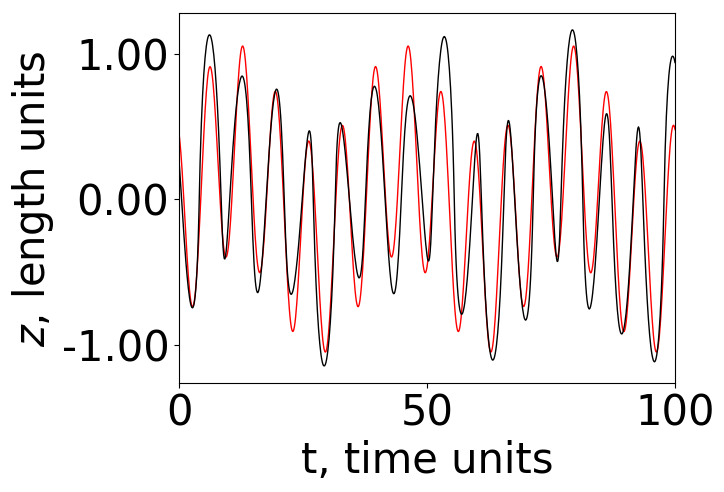}%
\end{minipage}
\begin{minipage}{.23\textwidth}
\centering
\includegraphics[width=\textwidth]{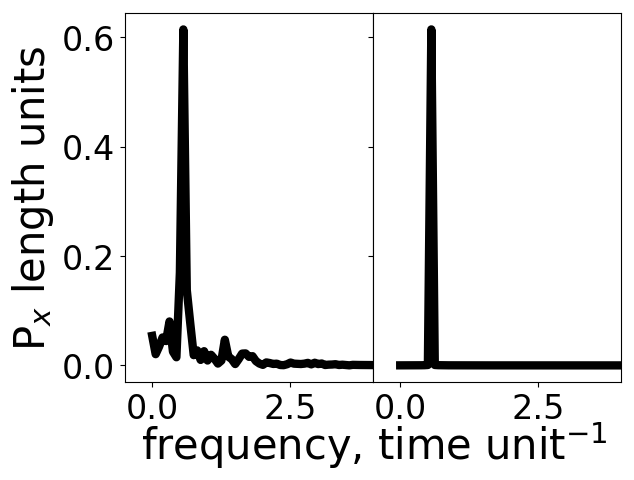}%
\end{minipage}
\begin{minipage}{.23\textwidth}
\centering
\includegraphics[width=\textwidth]{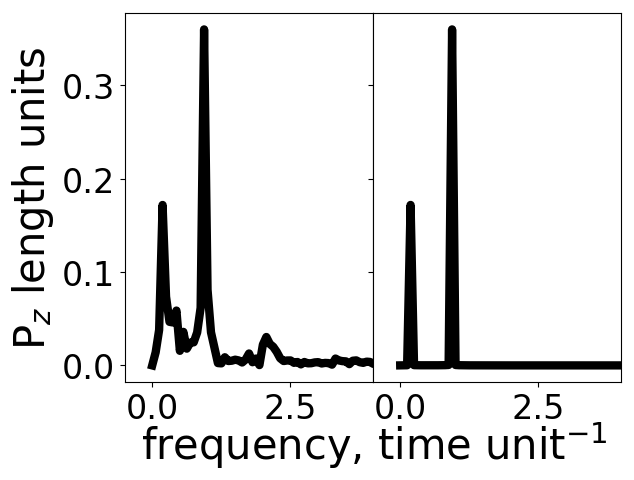}
\end{minipage}
\caption{Fitting of a `brezel'-like orbit from Fig.~\ref{fig:shrimp_brezel} by a ``perfect'' closed orbit. \textit{Left pair}: actual time series (black line) and least squares approximation (red) of it by eq.~\ref{eq_ap:funcs} for $x$ and $z$ coordinates, respectively. \textit{Right pair}: corresponding periodograms $P_x$ and $P_z$ of actual times series (\textit{left subpanels}) and approximated time series (\textit{right subpanels}).}
\label{fig_ap:ideal_orbit}
\end{figure}

Unfortunately, the actual orbits are far from being closed in our evolving $N$-body potential even if the frequencies are commensurate with good accuracy ($\Delta(f_z/f_x) = 10^{-3}$). Therefore, they are not very illustrative. To adequately distinguish the shape of the ``parent'' closed orbit, we constructed an artificial orbit based on an actual orbit in such a way that corresponding coordinate periodograms $P_x$ and $P_z$ of a new orbit would consist only of dominant lines of the actual orbit. To construct such an orbit, we fit the corresponding time series $x(t)$ and $z(t)$ by the following functions:
\begin{equation}
\begin{cases}
x(t) = A_x \cos(2\pi f_x t+ \phi_x) \\
z(t)=A_{z1} \cos(2\pi f_{z1} t+ \phi_{z1}) + A_{z2}\cos(2\pi f_{z2} t+ \phi_{z2})). 
\end{cases}
\label{eq_ap:funcs}
\end{equation}
Here $f_x$, $f_{z1}$ are dominant frequencies and $f_{z2}$ the second highest line extracted from $P_x$ and $P_z$ periodograms, respectively. All other parameters are fitted using the least squares method. Fig.~\ref{fig_ap:ideal_orbit} shows the results of fitting for `brezel'-like orbit from Fig.~\ref{fig:shrimp_brezel}. We experimented with different sets of lines and obtained, that the realistic orbital pattern is reproduced if we account for one line for in-plane oscillations and two lines for oscillations in the vertical direction.

\label{lastpage}
\end{document}